\documentclass{article}
\usepackage{microtype}
\usepackage{graphicx}
\usepackage{subcaption}
\usepackage{booktabs} 
\usepackage{hyperref}
\usepackage{subcaption}

\usepackage[accepted]{icml2026}
\makeatletter

\renewcommand{\@pa}[1]{%
  \ifcsname the@affil#1\endcsname\else
    \ifcsname @icmlsymbol#1\endcsname\else
      \stepcounter{@affiliationcounter}%
      \newcounter{@affil#1}%
      \setcounter{@affil#1}{\value{@affiliationcounter}}%
    \fi
  \fi
}

\renewcommand{\printAffiliationsAndNotice}[1]{\global\icml@noticeprintedtrue%
  {\let\thefootnote\relax\footnotetext{\hspace*{-\footnotesep}\ificmlshowauthors #1\fi%
      \ifdefined\icmlcorrespondingauthor@text
        Correspondence to: \icmlcorrespondingauthor@text.%
      \fi
      \ \\
      \Notice@String
    }%
  }%
}
\makeatother
\usepackage{amsmath}
\usepackage{amssymb}
\usepackage{mathtools}
\usepackage{amsthm}

\usepackage[capitalize,noabbrev]{cleveref}

\theoremstyle{plain}

\theoremstyle{definition}

\theoremstyle{remark}

\usetikzlibrary{matrix}

\begin{document}

\twocolumn[
  \icmltitle{What Does It Mean to Break a Distillation Defense?}

  \icmlsetsymbol{equal}{*}

  \begin{icmlauthorlist}
    \icmlauthor{Lena Libon}{yyy}
    \icmlauthor{Pura Peetathawatchai}{yyy}
    \icmlauthor{Michael Aerni}{yyy}
    \icmlauthor{Daniel Paleka}{yyy}
    \icmlauthor{Florian Tramèr}{yyy}

  \end{icmlauthorlist}

  \icmlcorrespondingauthor{Lena Libon}{llibon@ethz.ch}
  \vskip 0.08in
  {\centering ETH Zurich \par}

  \vskip 0.3in
]

\printAffiliationsAndNotice{}  

\begin{abstract}
Black-box LLMs (accessible only via API) are vulnerable to distillation attacks, in which an attacker queries the model and trains a student on its outputs. A recent line of work proposes \emph{output perturbation defenses} that modify the teacher's output to reduce student performance while preserving utility for legitimate users. As a relatively new family of approaches, output perturbation defenses lack a shared threat model, making it difficult to compare them, reason about composing them with other attacks, or evaluate their robustness against realistic adversaries. This underspecification matters beyond technical evaluation: when defenses are deployed to protect intellectual property or justify regulatory compliance, an imprecise threat model can create a false sense of security. We propose a threat model framework that describes attackers along three dimensions: a query budget, a data budget, and an interface profile that captures how attackers interact with the API. Using antidistillation sampling as a case study, we show that whether the defense is considered effective depends on the assumed threat model. We argue that future work on distillation defenses, along with any governance or policy frameworks built around them, should explicitly specify and stress-test attacker capabilities along our three dimensions.
\end{abstract}

\begin{figure*}[t]
    \centering
    \begin{tikzpicture}

    \def\pw{3.7}
    \def\ph{3.7}

    \begin{scope}[shift={(0,0)}]

        \fill[gray!5] (0,0) rectangle (\pw, \ph);
        \draw[gray!35, densely dashed, line width=0.8pt] (0,0) -- (\pw,\ph);

        \node[rotate=45, font=\footnotesize, text=black, anchor=center] 
            at (\pw*0.30, \ph*0.70) {Data-limited};
        \node[rotate=45, font=\footnotesize, text=black, anchor=center] 
            at (\pw*0.70, \ph*0.30) {Query-limited};

        \draw[-{Latex[length=2.5mm]}, thick] (0,0) -- (\pw+0.3,0);
        \draw[-{Latex[length=2.5mm]}, thick] (0,0) -- (0,\ph+0.3);

        \node[below, font=\small] at (\pw/2, -0.35) {Data budget};
        \node[rotate=90, above, font=\small] at (-0.35, \ph/2) {Query budget};

        \node[below, font=\footnotesize, text=gray!70!black] at (0, -0.05) {$0$};
        \node[below, font=\footnotesize, text=gray!70!black] at (\pw, -0.05) {$\infty$};
        \node[left, font=\footnotesize, text=gray!70!black] at (-0.05, 0) {$0$};
        \node[left, font=\footnotesize, text=gray!70!black] at (-0.05, \ph) {$\infty$};

        \node[font=\small\bfseries, above] at (\pw/2, \ph+0.35) {Query \& Data Budget};

    \end{scope}

    \node[font=\Large\bfseries, gray!40] at (\pw + 0.55, \ph/2) {$+$};

    \begin{scope}[shift={(\pw + 1.6, 0)}]

        \def\rw{8.1}
        \def\rh{\ph}

        \def\toph{1.95}
        \def\both{1.40}
        \def\gap{0.55}
        \pgfmathsetmacro{\topy}{\rh - \toph}
        \pgfmathsetmacro{\boty}{\topy - \gap - \both}

        \fill[gray!5, rounded corners=2pt] (0, \topy) rectangle (\rw, \rh);

        \node[font=\footnotesize\scshape, gray!55!black, rotate=90, anchor=south] 
            at (-0.15, \topy + \toph/2) {Generic};

        \tikzset{
            snode/.style={draw, rounded corners=1.5pt, inner sep=2pt, font=\tiny,
                          minimum width=0.7cm},
            anlabel/.style={font=\scriptsize\scshape, text=gray!55!black,
                            align=center},
            exlabel/.style={font=\tiny\itshape, text=gray!60!black, align=center},
        }

        \pgfmathsetmacro{\apicx}{\rw/2}
        \pgfmathsetmacro{\apicy}{\topy + \toph - 0.92}

        \def\armlen{2.7}

        \node[snode, fill=white]   (pG)   at (\apicx - \armlen, \apicy) {Input};
        \node[snode, fill=gray!20] (apiG) at (\apicx,           \apicy) {API};
        \node[snode, fill=white]   (rG)   at (\apicx + \armlen, \apicy) {Response};

        \draw[->, >=Latex, thin] (pG) -- (apiG);
        \draw[->, >=Latex, thin] (apiG) -- (rG);

        \node[anlabel, anchor=south] at (\apicx - \armlen, \apicy + 0.22) {Input channels};
        \node[anlabel, anchor=south] at (\apicx,           \apicy + 0.22) {Processing};
        \node[anlabel, anchor=south] at (\apicx + \armlen, \apicy + 0.22) {Output channels};

        \node[exlabel, anchor=north] at (\apicx - \armlen, \apicy - 0.22) 
            {e.g., prompt, prefill,\\sampling params};
        \node[exlabel, anchor=north] at (\apicx,           \apicy - 0.22) 
            {e.g., filtering,\\summarization};
        \node[exlabel, anchor=north] at (\apicx + \armlen, \apicy - 0.22) 
            {e.g., text,\\logprobs};

        \def\cellw{3.55}
        \def\cellh{\both}
        \def\cellgap{0.30}
        \def\arrstep{1.2}

        \pgfmathsetmacro{\gx}{(\rw - 2*\cellw - \cellgap) / 2}
        \pgfmathsetmacro{\gy}{\boty}

        \node[font=\footnotesize\scshape, gray!55!black, rotate=90, anchor=south] 
            at (-0.15, \gy + \cellh/2) {This work};

        \pgfmathsetmacro{\branchy}{\topy - 0.02}
        \pgfmathsetmacro{\splity}{\topy - \gap/2}
        \pgfmathsetmacro{\cellLcx}{\gx + \cellw/2}
        \pgfmathsetmacro{\cellRcx}{\gx + \cellw + \cellgap + \cellw/2}

        \draw[gray!45, densely dashed, line width=0.4pt]
            (\apicx, \branchy) -- (\apicx, \splity);
        \draw[gray!45, densely dashed, line width=0.4pt]
            (\cellLcx, \splity) -- (\cellRcx, \splity);
        \draw[gray!45, densely dashed, line width=0.4pt, -{Latex[length=1.5mm]}]
            (\cellLcx, \splity) -- (\cellLcx, \gy + \cellh + 0.02);
        \draw[gray!45, densely dashed, line width=0.4pt, -{Latex[length=1.5mm]}]
            (\cellRcx, \splity) -- (\cellRcx, \gy + \cellh + 0.02);

        \tikzset{
            snode/.style={draw, rounded corners=1.5pt, inner sep=2pt, font=\tiny,
                          minimum width=0.7cm},
        }

        \fill[gray!5, rounded corners=2pt]
            (\gx, \gy) rectangle +(\cellw, \cellh);
        \begin{scope}[shift={(\gx, \gy)}]
            \node[font=\scriptsize\bfseries, anchor=north west] at (0.12, \cellh - 0.1) {Black-box};
            \pgfmathsetmacro{\xmid}{\cellw/2}
            \node[snode, fill=white]   (p1)   at ({\xmid - \arrstep}, 0.42) {Prompt};
            \node[snode, fill=gray!20] (api1) at ({\xmid},            0.42) {API};
            \node[snode, fill=white]   (r1)   at ({\xmid + \arrstep}, 0.42) {Text};
            \draw[->, >=Latex, thin] (p1) -- (api1);
            \draw[->, >=Latex, thin] (api1) -- (r1);
        \end{scope}

        \fill[gray!5, rounded corners=2pt]
            ({\gx + \cellw + \cellgap}, \gy) rectangle +(\cellw, \cellh);
        \begin{scope}[shift={({\gx + \cellw + \cellgap}, \gy)}]
            \node[font=\scriptsize\bfseries, anchor=north west] at (0.12, \cellh - 0.1) {Prefill};
            \pgfmathsetmacro{\xmid}{\cellw/2}
            \node[snode, fill=white, minimum width=0.6cm]     (p3)   at ({\xmid - \arrstep}, 0.60) {Prompt};
            \node[snode, fill=purple!12, minimum width=0.6cm] (pf3)  at ({\xmid - \arrstep}, 0.26) {Prefix};
            \node[snode, fill=gray!20]   (api3) at ({\xmid},            0.43) {API};
            \node[snode, fill=white]     (r3)   at ({\xmid + \arrstep}, 0.43) {Text};
            \draw[->, >=Latex, thin] (p3.east) -- (api3);
            \draw[->, >=Latex, thin] (pf3.east) -- (api3);
            \draw[->, >=Latex, thin] (api3) -- (r3);
        \end{scope}

        \node[font=\small\bfseries, above] at (\rw/2, \rh + 0.35) {Interface Profile};

    \end{scope}

    \end{tikzpicture}
    \caption{\textbf{Our proposed threat model space for evaluating distillation defenses.} A complete threat model specifies a query budget, a data budget (left), and an interface profile (right). The interface profile is a collection of interface components spanning the input side (e.g., prefill), provider-side processing (e.g., output filtering, reasoning-trace summarization), and the output side (e.g., logprobs). In this paper we compare two interface profiles: a pure black-box API, and an API that permits response prefilling.}
    \label{fig:threat_model_new}
\end{figure*}
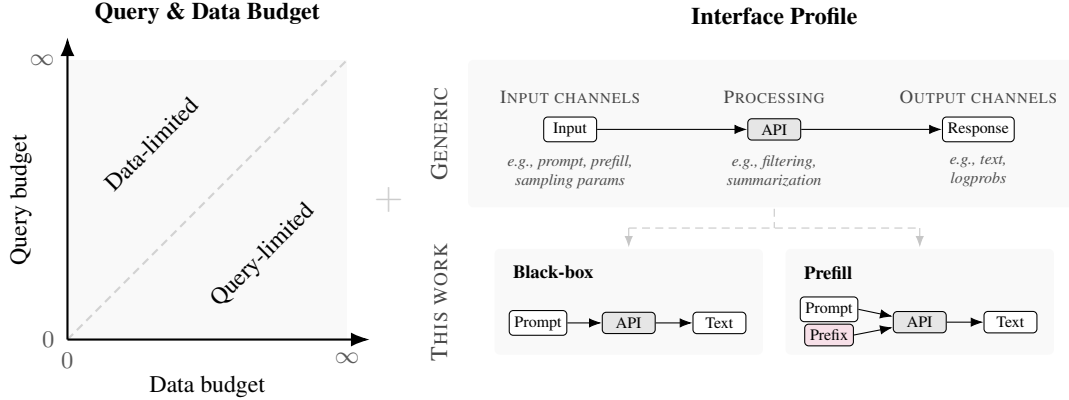

\section{Introduction}
Distillation is a process by which a \emph{student} model replicates the capabilities of a more powerful \emph{teacher} model by training on its outputs~\citep{hinton2015distilling, taori2023alpaca, peng2023instruction}, enabling smaller, cheaper models to achieve strong performance without the compute cost of training from scratch. As large language models have become widely accessible via APIs, however, distillation has also become a vector for extracting proprietary capabilities without authorization~\citep{anthropic2026distillation, openai2026memo}. From the perspective of frontier model providers, APIs that return rich outputs, including reasoning traces, token probabilities, or detailed generations, represent a potential forfeiture of intellectual property (IP), allowing competitors to replicate a model's capabilities without bearing the development costs. These concerns are reflected in provider terms of service and public communications~\citep{openai2025tos, GoogleCloud2026_GTIG_AI_Threat_Tracker}.

Existing defenses operate at three levels. At the \textit{system level}, providers implement rate limiting, coordinated activity monitoring, and legal enforcement through terms of service~\citep{anthropic2026distillation, GoogleCloud2026_GTIG_AI_Threat_Tracker, hulse2025deepseek}. These mechanisms operate outside the model and are independent of individual outputs. At the \emph{detection level}, statistical watermarks are embedded in outputs to enable post-hoc identification of distilled model content~\citep{kirchenbauer2023watermark, xu2026antidistillation}. More recent \emph{output perturbation} defenses \citep{jiang2026distillguard} modify model outputs to degrade student performance while preserving utility for legitimate users. Examples include fine-tuning the teacher~\citep{li2025doge}, rewriting reasoning traces after generation~\citep{ding2025information, ma2026protecting, hartman2026protecting}, and corrupting the output distribution at inference time, either via a proxy model~\citep{savani2025antidistillation} or a learned logit transformation~\citep{fang2026towards}.

Existing defenses at all three levels lack a unified threat model, making it difficult to compare them, reason about composition, or evaluate robustness against realistic adversaries. The problem is most acute for output perturbation defenses, where even the basic question of whether a defense works is undefined without specifying attacker capabilities. Existing work evaluates only against a simple implicit baseline: the attacker queries each prompt exactly once, collects outputs, and trains a student without further intervention~\citep{savani2025antidistillation, fang2026towards, li2025doge, ding2025information, ma2026protecting, jiang2026distillguard, hartman2026protecting}. This ignores how defenses perform against adversaries who allocate queries strategically, apply local post-processing, or interact through channels beyond the plain prompt.

This pattern recurs throughout adversarial machine learning: defenses proposed and tested against fixed, non-adaptive attackers were later broken by stronger adaptive ones. Examples span adversarial-example defenses~\citep{athalye2018obfuscated, tramer2020adaptive}, protective perturbations for artists~\citep{honigadversarial}, and LLM jailbreak and prompt-injection defenses~\citep{nasr2025attacker}. Defenses against distillation are at an analogous early stage. Yet the immediate obstacle is not only the absence of a stronger attack but the absence of any shared way to specify how strong an adversary is assumed to be. 

Our contribution is a framework for reasoning about attacker strength that makes these dimensions explicit. We characterize threat models along two quantitative dimensions: the attacker's \emph{query budget} (the total number of API calls to the teacher) and \emph{data budget} (the number of distinct input prompts available). Additionally, an \emph{interface profile} consisting of multiple interface components captures the structure of the attacker's interaction with the model: available input channels, provider-side processing, and exposed output channels. Together, a query budget, a data budget, and an interface profile fully specify a threat model. Figure~\ref{fig:threat_model_new} visualizes this space.

We use antidistillation sampling (ADS)~\citep{savani2025antidistillation} as a concrete case-study defense. ADS modifies the teacher's sampling distribution at inference time to reduce the performance of a distilled student while preserving output utility for legitimate users. To study the role of the interface profile, we vary a single component: whether the attacker can inject a prefix into the model's response. By evaluating ADS under these different threat models, we show how different assumptions can change conclusions about a defense's effectiveness. Although we focus on ADS, the framework itself is defense-agnostic: for detection-level defenses (e.g., watermarking), the setup is identical, except the attacker must also evade detection, while system-level defenses can be captured through budget constraints. 

We also argue that the lack of a shared threat model has consequences that extend beyond technical evaluation. As output perturbation defenses move from research into production~\citep{kim2026claudeleak}, policymakers, legal teams, and IP holders~\citep{openai2026memo, anthropic2026distillation} may rely on them as meaningful protection against distillation. But without an explicit threat model, neither deployers nor external auditors can know whether a defense that passed its evaluation would survive a more capable adversary. This ambiguity is not merely an academic limitation: it creates false assurances that may shape IP enforcement strategies, regulatory frameworks, and investment in better defenses. By substituting the appearance of protection for its substance and crowding out more robust alternatives, a deployed defense that overstates its coverage may ultimately set back IP protection efforts more than no defense at all.

\section{Background and Related Work} \label{sec:ads}
\subsection{Knowledge Distillation}
Knowledge distillation is a model compression technique in which a smaller \emph{student} model is trained to mimic the behavior of a larger \emph{teacher} model~\citep{hinton2015distilling}. Rather than training the student from scratch on raw data, the student learns from the teacher's outputs, which carry richer information than hard labels alone. For instance, the teacher's full output distribution over tokens encodes relative similarities between classes that a one-hot label discards~\citep{hinton2015distilling}.
\newpage

In the context of large language models, distillation typically takes the form of supervised fine-tuning on teacher-generated text: input prompts are collected, the teacher generates responses, and the student is trained on the resulting pairs~\citep{taori2023alpaca, peng2023instruction}. When the teacher produces explicit reasoning traces, these provide an especially rich training signal, since the student learns not just what answer to produce but how to decompose the problem~\citep{savani2025antidistillation}. Models distilled from frontier reasoning teachers can achieve strong performance on math and coding benchmarks at a fraction of the training cost~\citep{deepseek2025r1}.

\subsection{Antidistillation Sampling}
We use antidistillation sampling (ADS)~\citep{savani2025antidistillation} as a running example throughout this paper. ADS is a decoding-time defense that perturbs the teacher model's next-token distribution to reduce distillation effectiveness while preserving output utility. Given a prompt and partial generation, ADS samples from a modified distribution that up-weights tokens estimated to increase the student's downstream loss, using a finite-difference approximation through a proxy student model. The perturbation strength $\lambda \geq 0$ controls the trade-off between teacher utility and distillation degradation. At $\lambda = 0$, ADS reduces to standard temperature sampling. We focus on the \emph{high-utility regime} ($\lambda \leq 0.107$), where the teacher retains at least 70\% accuracy on the GSM8K dataset \cite{cobbe2021gsm8k}. In this regime, ADS achieves approximately 40\% student accuracy degradation for a 20\% teacher accuracy drop, substantially outperforming temperature sampling. Further details about ADS are provided in Appendix~\ref{app:ads_details}.

\subsection{Output Perturbation Defenses in Deployed Systems} Output perturbation defenses are deployed by several major providers in practice. A recent source leak of Anthropic's Claude Code exposed two of them \citep{kim2026claudeleak}: The first causes the API to silently inject fabricated tool definitions into the system prompt of first-party CLI sessions. The second one summarizes the assistant's inter-tool-call reasoning. A similar trace-summarization strategy is adopted by OpenAI, which exposes only model-generated summaries of \emph{o}-series reasoning rather than the raw chain of thought \citep{openai_reasoning_summaries}. 

\subsection{Limitations of Existing Evaluations} 
\paragraph{A shared, narrow evaluation pattern.} Output perturbation distillation defenses have, to our knowledge, been evaluated under narrow attacker profiles. In our framework (cf. Section~\ref{sec:threat_model}), most evaluations implicitly assume a minimal interface profile: the attacker sends a prompt through a single input channel, no provider-side processing is applied beyond the defense itself, and only the output text is returned. Both the query and data budget are bounded by the size of the benchmark dataset, each prompt is queried exactly once, and the collected text is used to train a student via supervised fine-tuning without filtering or post-processing.

ADS~\citep{savani2025antidistillation}, DOGe~\citep{li2025doge}, PART~\citep{ding2025information}, \citet{ma2026protecting}, TraceGuard~\citep{hartman2026protecting}, and the benchmark study DistillGuard~\citep{jiang2026distillguard} all follow this pattern, differing only in the choice of benchmark, teacher model, and student model. \citet{fang2026towards} extends the interface profile to expose logprobs but otherwise keeps the same setup.

DistillGuard explicitly acknowledges that stronger adversaries could reduce defense effectiveness, but does not formalize the capabilities of such adversaries. Concurrent work by \citet{allouah2026distillation} also identifies non-adaptive attackers as a limitation of current evaluations. They cast distillation as a minimax game in which an adaptive student reweights traces by their estimated learning value, and find that antidistillation defenses are significantly less effective against this student than passive evaluation suggests. TraceGuard asks a different question, whether the defense can be recognized rather than whether it can be broken, but the recognizer it considers still operates within the minimal interface profile.

\paragraph{Conflating query and data budgets.}  Even within this minimal threat model, an attacker can allocate a fixed query budget strategically, querying some prompts multiple times at the cost of skipping others, or selectively targeting the most informative prompts. Separating the query budget from the data budget captures exactly this flexibility: the query budget bounds total API calls, the data budget bounds the pool of available prompts, and the allocation strategy between them is left to the adversary. This distinction is particularly important in the setting of output perturbation defenses, because the cost of acquiring new high-quality prompts to extract useful reasoning traces and the cost of additional API calls scale very differently. To our knowledge, this separation is not made in the broader LLM model extraction literature either, where the attacker is consistently characterized by a single query budget~\citep{zhao2025survey} with each available prompt queried exactly once.

\begin{figure*}[t]
    \centering
    \begin{tikzpicture}[
        every node/.style={font=\small},
        entity/.style={align=center, inner sep=0pt, minimum width=1.8cm},
        costarrow/.style={-{Latex[length=2mm]}, BrickRed!80, line width=0.7pt},
        freearrow/.style={-{Latex[length=2mm]}, OliveGreen!70!black, line width=0.7pt},
        blackarrow/.style={-{Latex[length=2mm]}, black, line width=0.7pt},
        lbl/.style={font=\footnotesize, midway},
    ]
    \coordinate (baseline) at (0, -0.85);
    
    \node[entity] (teacher) at (0,0) {%
        \includegraphics[height=1.8cm]{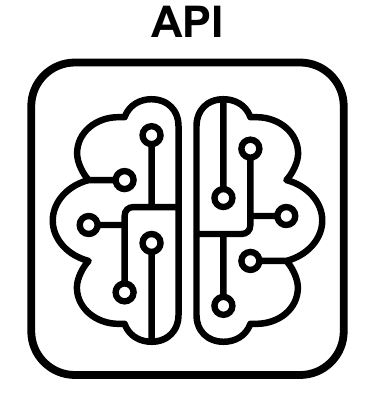}%
    };
    \node[font=\footnotesize\bfseries, anchor=north] (teacherlbl) at (teacher.south |- baseline) {Teacher API};
    
    \node[entity] (attacker) at (3.4,0) {%
        \includegraphics[height=1.2cm]{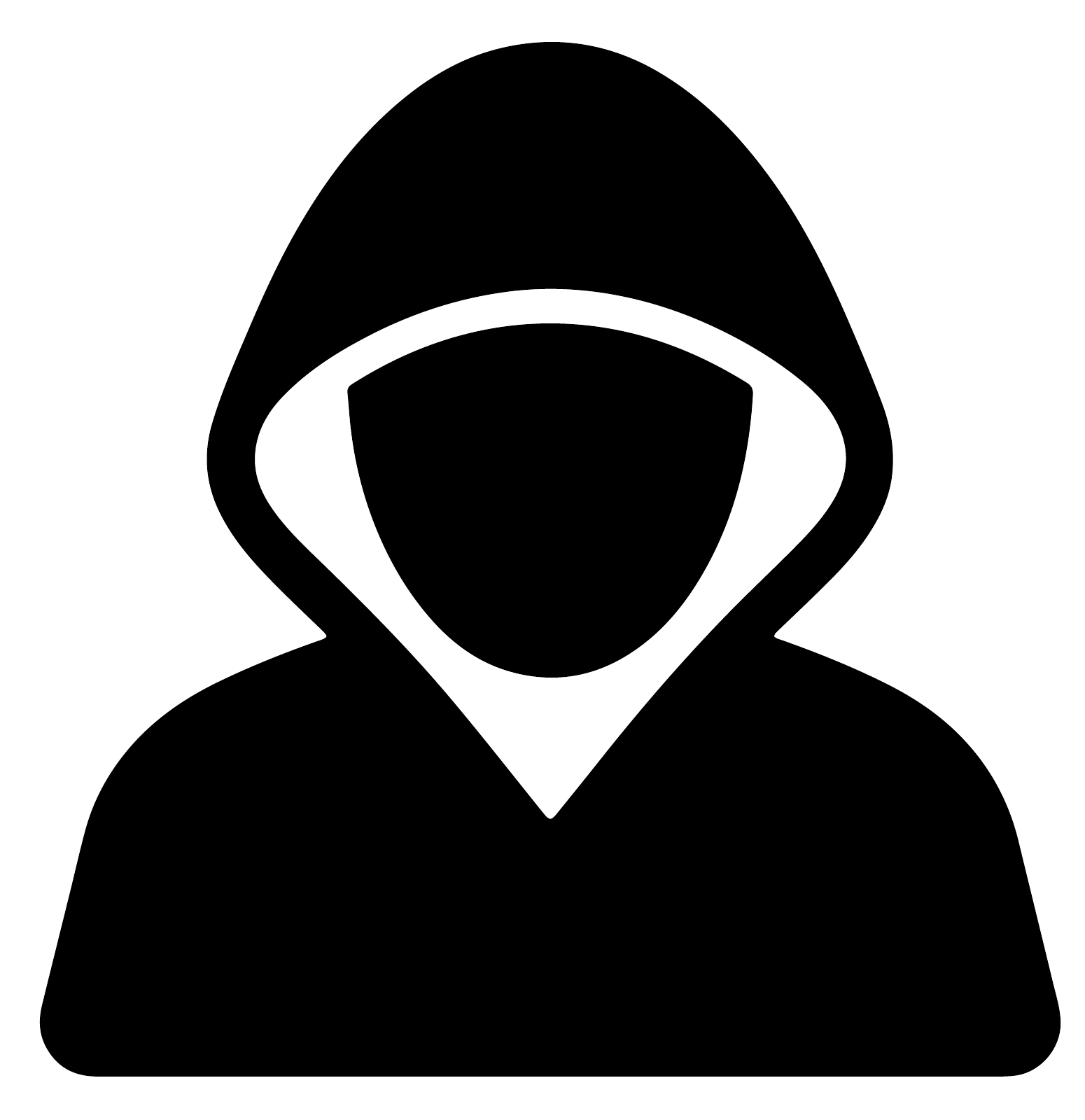}%
    };
    \node[font=\footnotesize\bfseries, anchor=north] (attackerlbl) at (attacker.south |- baseline) {Attacker};
    
    \node[entity] (data) at (6.8,0) {%
        \includegraphics[height=1.2cm]{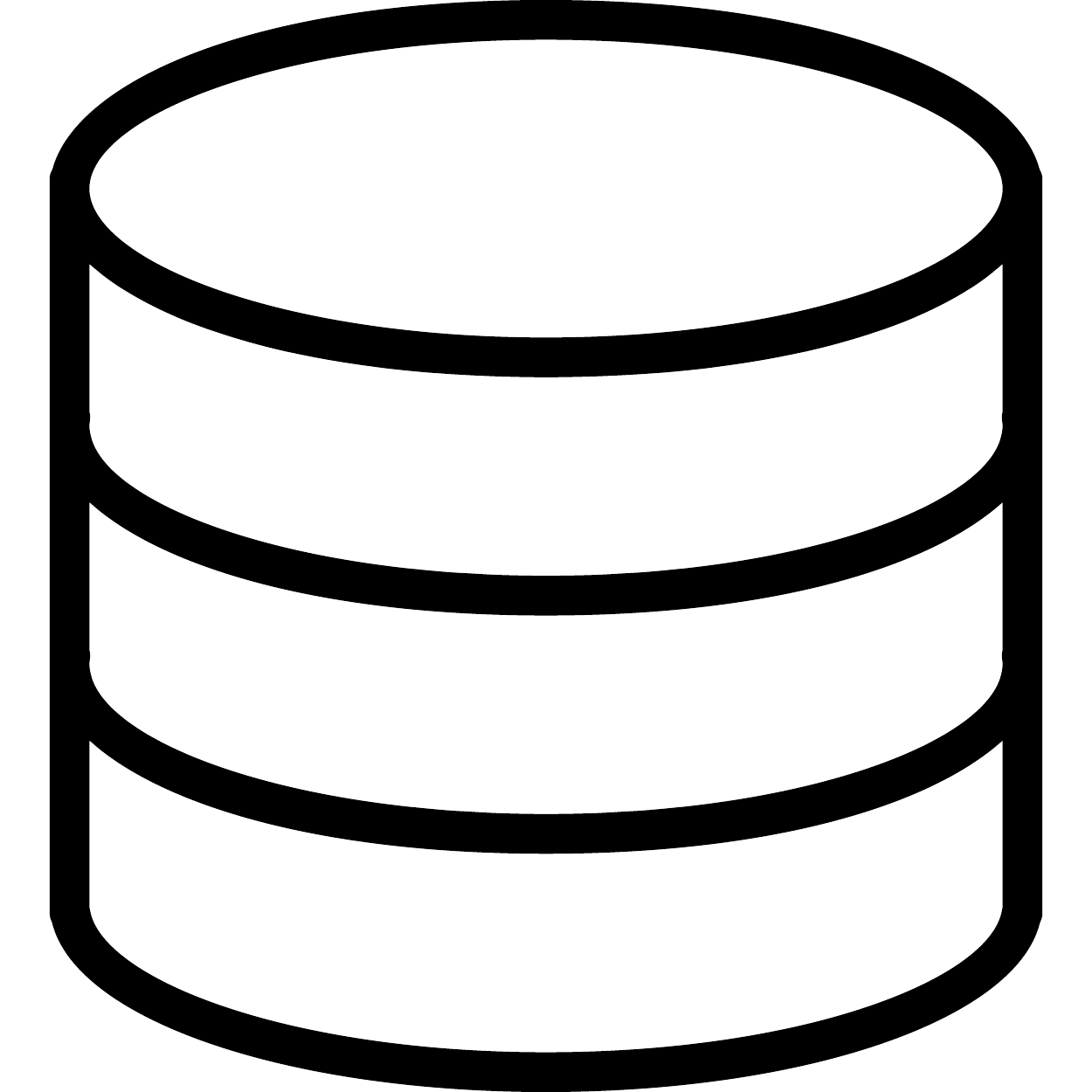}%
    };
    \node[font=\footnotesize\bfseries, anchor=north] (datalbl) at (data.south |- baseline) {Data};
    
    \node[entity] (student) at (10.2,0) {%
        \includegraphics[height=1.4cm]{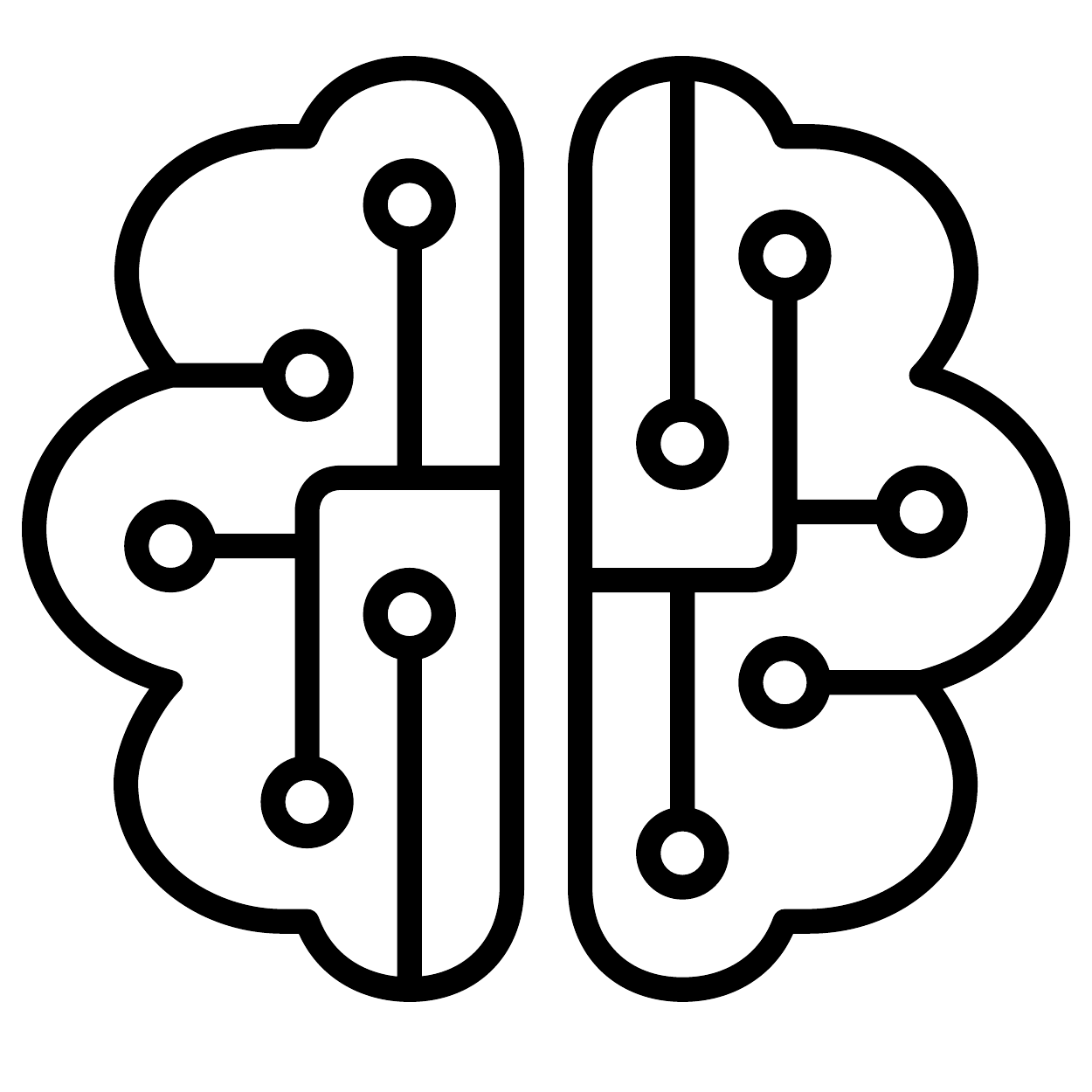}%
    };
    \node[font=\footnotesize\bfseries, anchor=north] (studentlbl) at (student.south |- baseline) {Distilled model};
    
    \draw[costarrow] ([yshift=2pt]attacker.west) -- node[lbl, above] {prompts} ([yshift=2pt]teacher.east);
    \draw[blackarrow] ([yshift=-5pt]teacher.east) -- node[lbl, below] {responses} ([yshift=-5pt]attacker.west);
    \draw[freearrow] (attacker.east) -- node[lbl, above] {responses} (data.west);
    \draw[freearrow] (data.east) -- node[lbl, above] {train} (student.west);
    
    \node[font=\footnotesize\bfseries, anchor=north] (captitle) at (5.1, -1.9) {Attacker's capabilities};
    
    \node[draw=BrickRed!50, fill=BrickRed!5, rounded corners=2pt, 
          inner sep=4pt, font=\footnotesize, anchor=north east,
          text width=4.8cm, align=left]
          (legend_costed) at (5.0, -2.45) {%
        \textcolor{BrickRed!80}{\textbf{Counted toward attacker budget}}\\[2pt]
        e.g.\ prompt (re-)querying
    };
    \node[draw=OliveGreen!50, fill=OliveGreen!5, rounded corners=2pt, 
          inner sep=4pt, font=\footnotesize, anchor=north west,
          text width=5.2cm, align=left]
          (legend_free) at (5.2, -2.45) {%
        \textcolor{OliveGreen!70!black}{\textbf{Free (local compute)}}\\[2pt]
        e.g.\ post-processing, student training
    };
    
    \end{tikzpicture}
    \caption{\textbf{Distillation attack pipeline and attacker budget model.} The attacker queries the teacher API and collects responses to build a training dataset. Only teacher API queries and input prompts count toward the attacker's budget (\textcolor{BrickRed}{red}); all local computation, including filtering, post-processing, and student training, is free (\textcolor{OliveGreen}{green}).}
    \label{fig:cost_model}
\end{figure*}
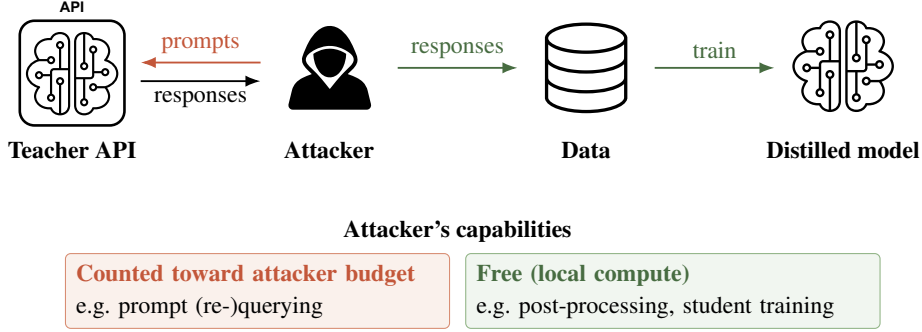

\section{Threat Model Framework} \label{sec:threat_model}

A defense against a distillation attack is only meaningful relative to a concrete attacker. We propose a framework (see Figure~\ref{fig:threat_model_new}) that characterizes threat models as tuples $(Q, D, P)$ where $Q \in \mathbb{N}$ is a \emph{query budget}, $D \in \mathbb{N}$ is a \emph{data budget}, and $P$ is an \emph{interface profile}: a structured, qualitative description of how the attacker can observe and interact with the API.

\subsection{Query and Data Budgets}

The distillation pipeline and our budget model are illustrated in Figure~\ref{fig:cost_model}. Only teacher API queries and input prompts constitute the attacker's cost. All other computation, such as student training, filtering low-quality generations, or post-processing outputs, is treated as free. This modeling choice reflects how API access is governed by the provider,
whereas local computation scales with the attacker's own resources.

Within this cost model, we distinguish two dimensions: (1) the \emph{query budget}~$Q$, the total number of API calls the attacker can make, and (2) the \emph{data budget}~$D$, the number of distinct input prompts available to the attacker. When $Q > D$, the attacker can obtain multiple generations per prompt, enabling strategies such as requerying prompts that produced incorrect outputs. When $Q < D$, the attacker cannot query every available prompt and must prioritize a subset of the data. They may query many prompts once, query a smaller subset multiple times, or allocate queries adaptively. In either case, the attacker is free to allocate queries unevenly between prompts. These separate budgets thus capture attacker strategies that a single dataset size parameter would conflate.

\subsection{Interface Profile}
The interface profile $P$ captures the structure of the attacker's interaction with the API, orthogonal to the number of queries or prompts used. We decompose $P$ into three stages, illustrated in Figure~\ref{fig:threat_model_new} (right): the \emph{input channels} available to the attacker, the \emph{provider-side processing} applied between the model's post-defense output and the response returned to the attacker, and the \emph{output channels} through which information is exposed. Each stage is a set of \emph{interface components}, and each component is a discrete design choice made independently by the provider. A defense that is robust against one interface profile may collapse once a single component is changed, so specifying $P$ is essential for any claim about a defense's effectiveness.

On the \emph{input side}, the most basic channel is the prompt itself, but APIs commonly admit additional signals like a hidden or attacker-controlled system prompt, sampling parameters (e.g., temperature, top-$p$, top-$k$, repetition penalty, and stop sequences), a response prefill forcing the model to begin its answer with a chosen prefix, or a multi-turn conversational state that allows the attacker to build context across calls.

For \emph{provider-side processing}, the provider may apply pipeline steps between the post-defense model output and the response the attacker sees. Examples include safety and copyright filters that redact or replace portions of the output, truncation of responses that exceed a maximum length, and reasoning-trace summarization. Some of these steps can also be counted as distillation defenses. For instance, summarizing reasoning traces is used both to reduce latency and to limit reasoning extraction. From a threat-modeling perspective, what matters is the transformation's effect on the attacker rather than the provider's intent.

On the \emph{output side}, the generated text is the baseline signal, but providers can additionally return token-level logprobs (over the full vocabulary or a top-$k$ truncation) or intermediate tool calls.

\subsection{Mapping to Practice}
Each dimension of our framework corresponds to concrete constraints that attackers face and providers impose. In previously observed attacks, carefully designed prompts have been used to elicit high-quality responses targeting specific capabilities, or to create broad task distributions for training \citep{anthropic2026distillation}.

However, producing such prompts at scale is costly, so an attacker cannot simply acquire new prompts on demand. The data budget reflects this difficulty in obtaining suitable prompts. Meanwhile, the query budget is bounded by the attacker's financial resources and the inference costs, by provider-deployed mechanisms at the system level (usage quotas, rate limits, behavioral fingerprinting, coordinated activity monitoring, and terms-of-service enforcement) \citep{anthropic2026distillation, GoogleCloud2026_GTIG_AI_Threat_Tracker, hulse2025deepseek}, as well as by the attacker's capacity to circumvent them. For instance, OpenAI alleges that DeepSeek developed programmatic access patterns and routed queries through obfuscated third-party routers to evade detection \citep{openai2026memo}. 

The interface profile is determined by the provider's API and varies across companies and model generations. Input channels differ: response prefill is supported by some providers like DeepSeek \citep{deepseek_chat_prefix_completion} and previously by earlier Claude models \citep{anthropic_claude_putting_words_in_mouth} but was removed from Claude Opus~4.6 and Sonnet~4.6 onwards \citep{anthropic_claude_prefill_response}. Provider-side processing also varies: OpenAI summarizes reasoning traces for its \emph{o}-series models \citep{openai_reasoning_summaries}, Anthropic does the same for Claude $4$ \citep{vercel_reasoning_summarized_vs_full_thinking}, and all major providers apply content filtering. Output channels have changed over time, with logprobs being the most visible example, since they were widely available early on, but have been restricted by many providers due to extraction risks \citep{carlini2024stealing}. Each threat model in our framework thus corresponds to a realistic deployment scenario defined by a specific provider's API at a specific point in time.

\section{Case Study: Antidistillation Sampling}\label{sec:attacks}

To demonstrate the value of our framework, we present a case study on antidistillation sampling (ADS) that shows how varying the assumed threat model can change conclusions about a defense's effectiveness. For direct comparison with \citet{savani2025antidistillation}, we match their dataset, models, and training setup throughout.

Our goal is not to find the strongest possible attack against ADS, but to show that even small, well-motivated changes to the threat model are sufficient to reverse conclusions about whether a defense works. This is precisely the risk that an underspecified threat model creates. Rather than sweeping the full threat model space, which would require a separate study for each combination of interface components, budget regimes, and student architectures, we therefore vary a small number of dimensions while holding others fixed. We fix every interface component except one: whether the attacker has prefill access.

We focus on prefill for two reasons. First, it is a design choice that some commercial APIs expose while others have explicitly removed. Second, unlike richer output channels such as logprobs, which would require changing the student training procedure to use soft targets, prefill can be studied within the exact evaluation setup of the ADS paper. This isolates the effect of a single interface component and shows that conclusions about defense effectiveness can depend on choices often treated as implementation details.

We conclude the section with a discussion of threat model variations we do not instantiate.

\paragraph{Experimental setup.} Throughout this section, we fix the data budget to a 70\% subset of GSM8K \cite{cobbe2021gsm8k} ($D = 5{,}231$ samples), matching the original ADS evaluation 
\citep{savani2025antidistillation}. We adopt the same teacher (DeepSeek-R1-Distill-Qwen-7B), proxy student (Qwen2.5-3B), and student model (Llama-3.2-3B), and use the same training setup, including fine-tuning the student via LoRA with identical hyperparameters. Our threat model is therefore the tuple $(Q, D, P)$ with $D=5{,}231$ fixed, $Q$ varied across experiments, and $P$ taking one of the two values: a restrictive black-box interface profile (text-only output, no additional input channels beyond the prompt, no provider-side processing), or the same profile extended to admit a response prefill on the input side. 

\paragraph{Evaluation.} We define the high-utility regime as the settings where the teacher retains at least 70\% accuracy on GSM8K ($\lambda \leq 0.107$), and evaluate five values $\lambda \in \{0.0553, 0.0632, 0.0711, 0.0868, 0.107\}$ within this range. See Appendix~\ref{app:extended_results} for lower-utility results. To measure student performance, we follow \citet{savani2025antidistillation} and use answer forcing (AF): after free-form generation from the prompt, we concatenate ``\texttt{\textbackslash n\textbackslash n**Final Answer**\textbackslash n[\textbackslash boxed\{}" and generate 32 additional answer tokens. Appendix~\ref{app:example_traces} provides examples of raw and ADS-attacked traces.

\subsection{Weak-Attacker Model}

We begin with the weakest attacker: black-box interface profile, with query budget equal to data budget ($Q = D = 5{,}231$). This is the threat model under which ADS was originally evaluated. Even here, since our framework treats any computation the attacker performs on collected outputs as free, a realistic attacker has access to free local post-processing that the defense cannot prevent.

\begin{figure}[t]
    \centering
    \includegraphics[width=0.95\linewidth]{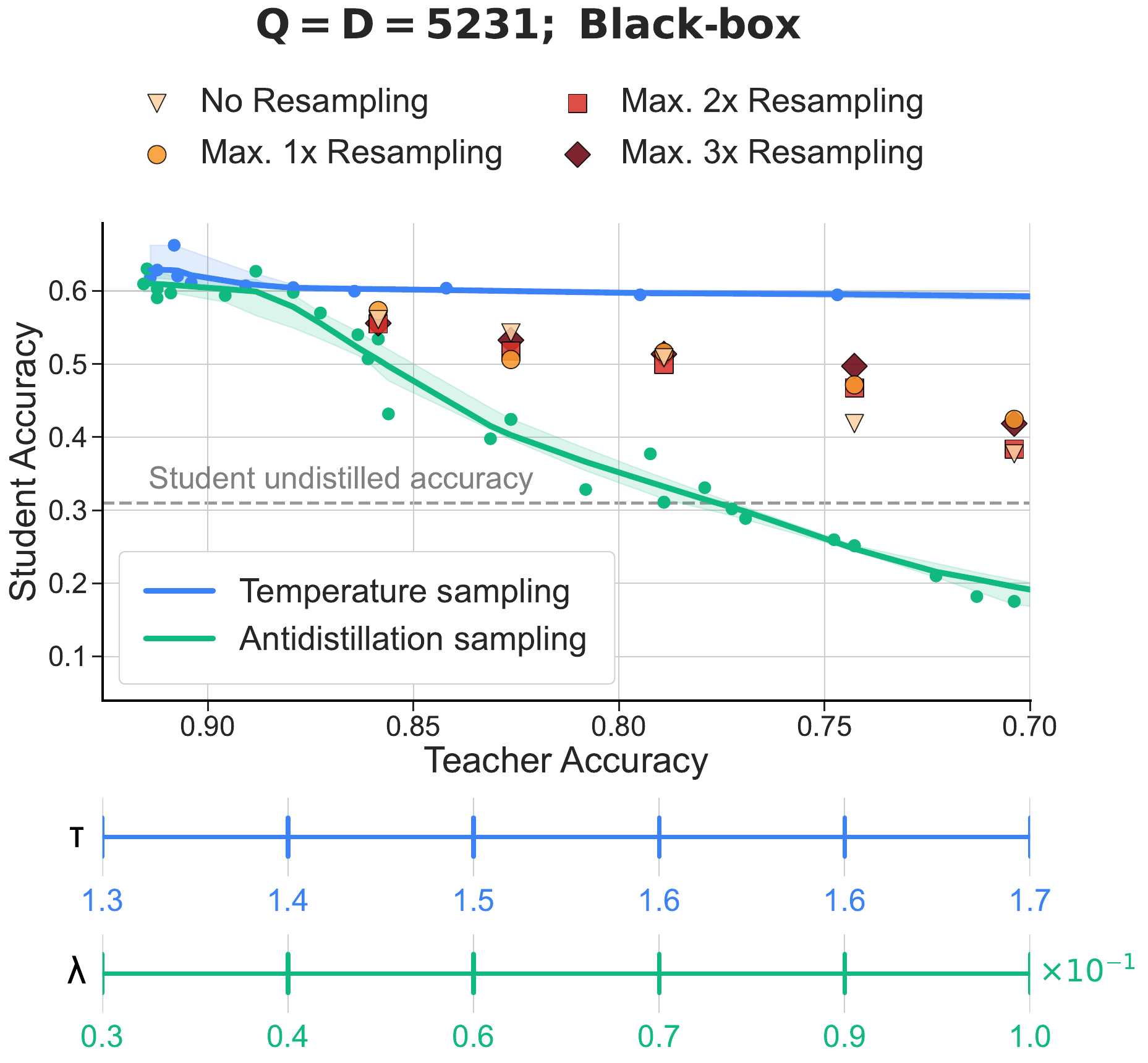}
    \caption{\textbf{Simple post-processing already significantly recovers student accuracy.}     Student accuracy on GSM8K under max-$k$ resampling ($k \in \{1,2,3\}$) and no resampling, all with repetition deletion applied, within ADS's own threat model ($Q = D = 5231$; black-box access). All strategies significantly outperform the originally reported ADS baseline, and differences between resampling strategies are small.}
    \label{fig:query_allocation}
\end{figure}

ADS outputs often contain repetitive passages that are straightforward to detect and remove. Deleting these across all collected traces before training requires no additional queries and no knowledge of the defense mechanism. As shown by the ``No Resampling'' points of Figure~\ref{fig:query_allocation}, this single step already lifts student accuracy significantly above the originally reported baseline across the high-utility regime. Because this post-processing is within the capabilities of any attacker operating under ADS's own assumptions, the utility gap reported by \citet{savani2025antidistillation} overestimates the defense's effectiveness even in the original threat model.

We additionally test query reallocation, a second strategy available within the same interface profile. With a fixed query budget, an attacker can requery prompts that produced incorrect answers rather than querying every prompt exactly once. We experiment with max-$k$ resampling: a prompt that produces an incorrect answer is requeried up to $k$ additional times, yielding at most $k+1$ generations per prompt in total, for $k \in \{1,2,3\}$. This trades prompt coverage for per-prompt correctness. As Figure~\ref{fig:query_allocation} shows, all $k$ values perform similarly to $k = 0$ once repetition deletion is applied, so the gain over the reported ADS baseline is driven primarily by post-processing rather than by resampling.

The takeaway is that the originally reported baseline corresponds to an attacker who performs no local post-processing, and that a slightly stronger attacker who simply deletes repetitions, still operating within ADS's own threat model, already recovers much of the reported utility gap. Defense evaluations that ignore free local post-processing risk overstating robustness. Because local computation is outside the provider's control, any realistic attacker will exploit it, and the most a defense can do is acknowledge that its effective utility gap is smaller than reported. More broadly, we echo the long-standing recommendation in the adversarial ML and privacy literature that defenses should be evaluated against adaptive adversaries rather than fixed baselines~\citep{carlini2019evaluating, honigadversarial, qievaluating, nasr2025attacker, aerni2024evaluations}. For output perturbation distillation defenses, this means stress-testing against attackers who post-process outputs, allocate queries strategically, and exploit available interface components.

\begin{figure*}
    \centering
    \begin{minipage}[t]{0.48\linewidth}
        \centering
        \includegraphics[width=\linewidth]{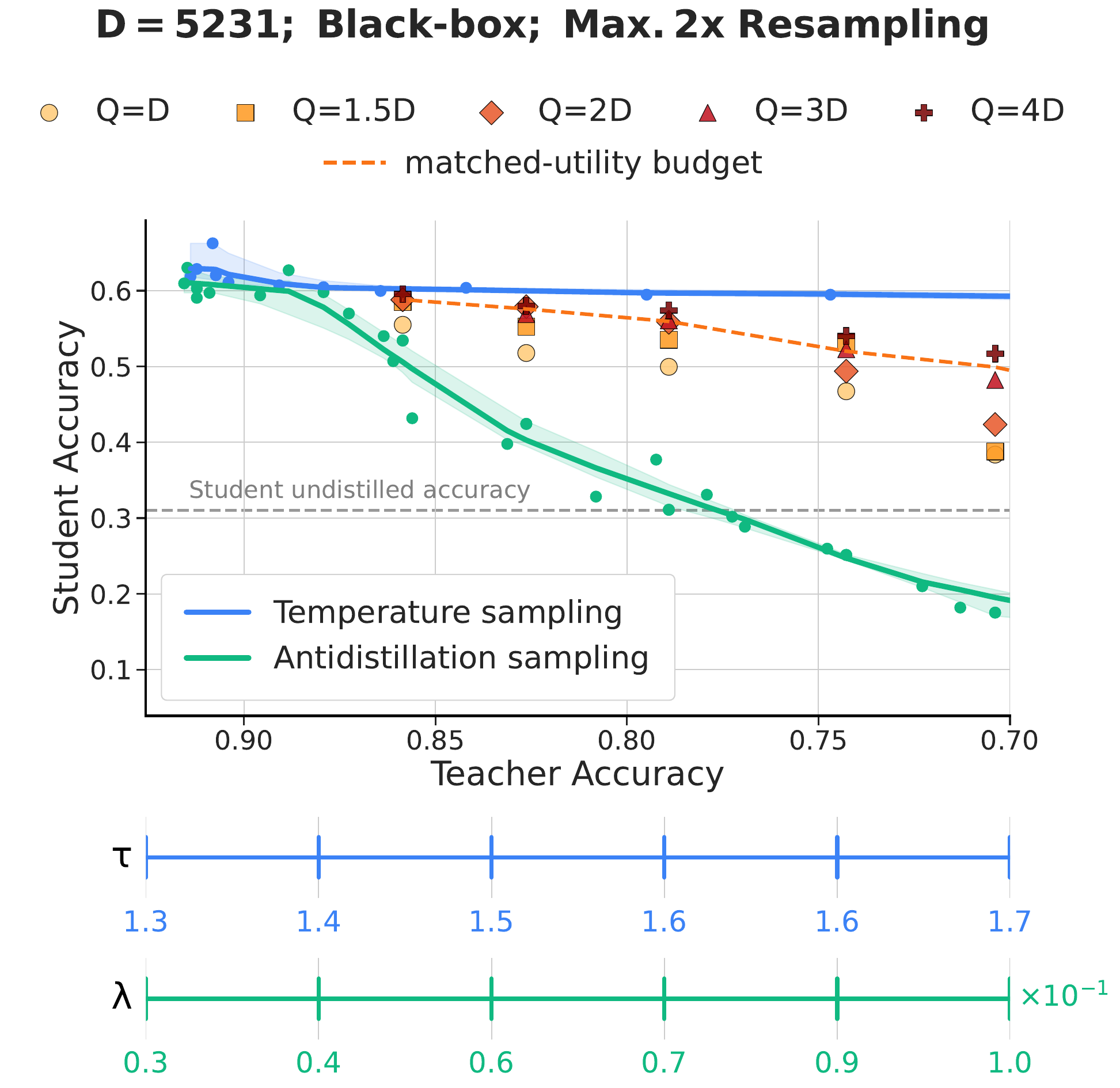}
        \captionof{figure}{\textbf{Increasing the query budget largely neutralizes ADS.} Student accuracy on GSM8K under max-2 resampling with repetition deletion, for query budgets $Q \in \{D, 1.5D, 2D, 3D, 4D\}$, under black-box access. The matched-utility budget curve is interpolated from the existing discrete points rather than measured directly. For the three smallest $\lambda$ values, $Q \geq 2D$ almost recovers the student accuracy achieved under temperature sampling. Across higher $\lambda$ values the matched-utility budget achieves near-complete recovery, outperforming several fixed budget settings.}
        \label{fig:budget_allocation_1}
    \end{minipage}
    \hfill
    \begin{minipage}[t]{0.48\linewidth}
        \centering
        \includegraphics[width=\linewidth]{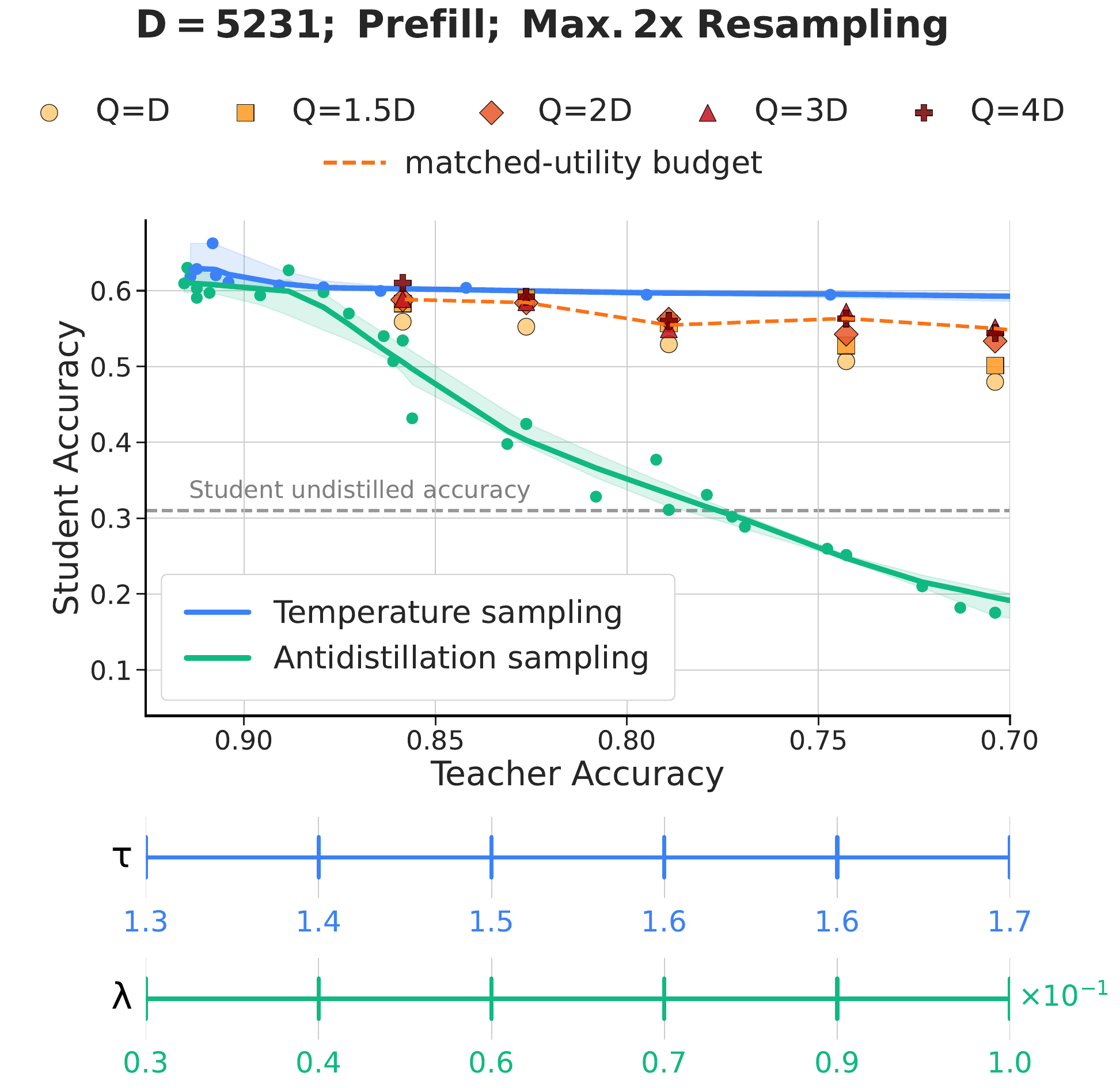}
        \captionof{figure}{\textbf{A single-token prefill neutralizes ADS across the entire high-utility regime.} Student accuracy on GSM8K under max-$2$ resampling with repetition deletion and a one-token ``First'' prefix applied to every generation, for varying query budgets $Q \in \{D, 1.5D, 2D, 3D, 4D\}$, under the prefill access profile. At higher budgets, student accuracy almost fully recovers that of temperature sampling across all $\lambda$ values. The matched-utility budget curve is also shown, also achieving near-complete recovery.}
        \label{fig:budget_allocation_first}
    \end{minipage}
\end{figure*}
\vspace{-0.2em}
\subsection{Scaling the Query Budget} \label{sec:cost_analysis}
We next relax the constraint $Q = D$ and examine how student accuracy scales as the query budget grows 
beyond the data budget. We vary 
$Q \in \{D,\, 1.5D,\, 2D,\, 3D,\, 4D\}$, use max-$2$ resampling throughout, and apply repetition deletion to all collected outputs. When the attacker exhausts the dataset with query budget remaining, resampling restarts from the beginning.

This threat model is more realistic than the $Q = D$ setting, as API providers impose query limits that are independent of how many distinct prompts an attacker has assembled. Varying the query budget also reflects the fact that an attacker who invests additional effort, whether by circumventing rate limits, paying for higher-tier access, or distributing queries across multiple accounts, obtains a larger effective query budget while operating on the same fixed pool of prompts.

Figure~\ref{fig:budget_allocation_1} shows that student accuracy increases consistently with $Q$, and that for the three smallest perturbation $\lambda$ values, a budget of $Q \geq 2D$ almost fully recovers the student accuracy under temperature sampling. An attacker who can increase their query budget can therefore largely circumvent the effectiveness of ADS across most of the high-utility regime. The quantitative conclusion about how effective ADS is depends directly on what value of $Q$ we assume. Tying $Q$ to $D$ has no particular justification, as the two are independent parameters and are rarely equal in practice.

This is a concrete instance of the broader problem: two threat models differing only in a single scalar parameter can yield conclusions as far apart as 40\% student accuracy degradation versus only a few percent. Without specifying $Q$ as part of the evaluation, a reported utility gap is not a property of the defense, but rather of the threat model used to measure it.
\vspace{-0.2em}
\paragraph{Matched-utility budget.} The previous subsection swept $Q$ as an external assumption, leaving open which value is the right one to evaluate at. The matched-utility budget instead derives $Q$ from what the defense's degradation already costs a legitimate user.

ADS degrades output quality for all queries to the API. A legitimate user who needs correct answers will therefore issue more than one query per prompt on average, retrying prompts that produced incorrect outputs. An attacker operating under the same interface has access to the same degraded outputs and has no reason to operate under a stricter query budget than a legitimate user.

A natural calibration is therefore to set $Q$ equal to $D$ multiplied by the expected number of queries a legitimate user would need to obtain a correct answer for each prompt. We call this the \emph{matched-utility budget}. This budget follows directly from ADS's own utility cost: the stronger the perturbation, the higher the legitimate user's expected query count, and the larger the matched-utility budget the attacker can justify. 

Figure~\ref{fig:budget_allocation_1} shows student accuracy under the matched-utility budget across the high-utility regime. Notably, our estimate of the expected query count is computed only over prompts solved within $16$ attempts (over 90\% of prompts in the high-utility regime), so the true matched-utility budget is in fact larger. Across higher $\lambda$ values, where ADS imposes the greatest utility cost, the matched-utility budget achieves near-complete recovery of temperature sampling performance, without requiring the attacker to exceed what a legitimate user would spend. In contrast, matched-utility resampling has a negligible effect under temperature sampling, where the resampled and standard curves are almost identical (see Appendix~\ref{app:resampling_temperature_buget}).

\subsection{Interface Profile: Enabling Prefill Access}\label{sec:prefill}

We now turn from budget dimensions to the interface profile. We extend the input side by a single component, \emph{prefill access}, while keeping provider-side processing and output channels unchanged.

Concretely, we prepend the single token ``\texttt{First}'' to every generation. This requires no additional queries and no changes to the training pipeline. All other experimental choices match Section~\ref{sec:cost_analysis}, including max-$2$ resampling, repetition deletion, and $Q \in \{D, 1.5D, 2D, 3D, 4D\}$.

Figure~\ref{fig:budget_allocation_first} shows that this one-token prefill improves student accuracy by an additional $\approx$5 percentage points compared to the black-box setting (Figure~\ref{fig:budget_allocation_1}) across all values of $Q$. At higher budgets, student accuracy almost fully recovers that of temperature sampling across the entire high-utility regime. The reason for this improvement is that ADS perturbs the teacher's next-token distribution toward tokens estimated to harm the student, but once the model is conditioned on even a single plausible prefix token, the remaining output becomes significantly less noisy, leading to more correct answers and near-baseline student accuracy. We provide further analysis in Appendix~\ref{sec:prefill_access_appendix}.

\vspace{1em}

The broader implication is that a defense whose robustness collapses under a one-token prefix is only meaningful relative to an interface profile in which prefill is unavailable. Several currently deployed APIs expose prefill~\citep{deepseek_chat_prefix_completion}, and the feature was available on Claude before being explicitly disabled on recent models~\citep{anthropic_claude_putting_words_in_mouth}, suggesting that providers themselves recognize the interface profile as a security-relevant design choice. A defense evaluated without specifying its assumed interface profile leaves practitioners unable to judge whether the evaluation applies to the API they are actually deploying behind.

\subsection{Other Threat Models}

Our case study spans two regions in the threat model space: varying the query budget $Q \in \{D, \dots, 4D\}$ and changing a single component of $P$ (prefill on the input side) at a fixed data budget $D$. Our goal is not to find the strongest possible attack against ADS, but to show that the assumed threat model determines conclusions about its security properties. The two regions are sufficient to demonstrate this. Here, we briefly discuss the remaining regions, each representing an additional axis along which a defense's reported effectiveness could shift.

\paragraph{Data budget regimes.} The data budget is fixed in our evaluation because \citet{savani2025antidistillation} fix it to 70\% of GSM8K. Varying $D$ would require a different dataset or a way to synthesize additional GSM8K-like prompts. We also do not evaluate $Q < D$ scenarios, in which the attacker must strategically select a subset of prompts to query. In such a setting, active selection strategies become relevant and could favor the attacker beyond uniform sampling.

\paragraph{Other input channels.} Our prefill experiment modifies a single input component. Other input-side components, such as sampling parameters, multi-turn conversational state and access to tools, would each open additional possibilities for an attacker to steer generation toward more distillation-useful outputs. The effect we observe from a one-token prefix suggests that any input channel that reduces the entropy of the output distribution will similarly decrease ADS's perturbation effects.

\paragraph{Provider-side processing.} We kept the provider's pipeline fixed throughout, corresponding to an interface profile with no filtering, truncation, summarization, or redaction between ADS and the attacker. Deploying ADS behind an API that summarizes reasoning traces, applies safety filters that redact parts of the output, or truncates long generations would change what the attacker observes per query, and could interact with ADS's perturbation. 

\paragraph{Output channels.} We do not evaluate logprob-exposing interface profiles. Few deployed models currently expose them (cf. Section~\ref{sec:threat_model}), and prior work already shows that token-level logprobs carry more signal than sampled tokens alone \citep{carlini2024stealing}.

Taken together, the regions we cover and those we do not point to the same conclusion: each threat-model dimension, budget or interface, carries enough degrees of freedom to change quantitative outcomes, and in some cases qualitative ones. A defense evaluation pinned to a single point in this space says very little about robustness elsewhere. 

\section{Discussion}
\paragraph{Fully exploiting the threat model matters.}
Our experiments show that even within a single fixed threat model, conclusions about a defense depend on how fully the attacker exploits it. Under the threat model of the original ADS paper, student accuracy for higher $\lambda$ values falls below the undistilled baseline, suggesting the perturbed traces are useless and an attacker would not use them. However, repetition deletion alone requires no additional queries and no extra interface components; it thus operates fully within ADS's own threat model, yet lifts student accuracy above the undistilled baseline for all $\lambda$ in the high-utility regime.
Fully exhausting the attacker's power within the assumed threat model is thus necessary to make any robustness claim. Concurrent work by \citet{allouah2026distillation} reaches the same conclusion: reweighting collected traces by their estimated learning value, which our framework treats as free local post-processing, lets an adaptive student substantially outperform a passive one within the same threat model.

\vspace{-0.5em}

\paragraph{Threat models as an evaluation standard.} Changing the threat model alters the conclusions about a defense's effectiveness. Scaling the query budget recovers most of the utility gap for the smaller $\lambda$ values, and granting the attacker a single additional input channel, prefill, nearly recovers temperature-sampling performance across the high-utility regime. The same defense appears effective or ineffective depending on how the attacker uses their budget and which components of the interface profile the provider exposes.

This variability is not a property of ADS specifically, but rather what happens to any output perturbation defense if the attacker's power is left unspecified. A defense described without a threat model cannot be meaningfully compared to another, because the two may be measured against different implicit adversaries. Nor can it be meaningfully deployed: providers may rely on it under attacker assumptions stronger than those under which it was evaluated.

Our results in Section \ref{sec:prefill} illustrate this directly. A provider who deploys ADS behind an API that exposes prefill would be offering near-complete recovery of student accuracy to any attacker willing to prepend a single token, while believing that they are protected. This is worse than having no defense at all, as it creates a false sense of security that may displace the search for real security. Making the threat model an explicit part of any robustness claim is therefore a precondition for both scientific comparison and responsible deployment. We argue it should be adopted as a minimum reporting standard for output perturbation defense research.

Distillation attacks are no longer just an academic concern. Major providers frame them as both a form of intellectual property theft and a national-security risk, since illicitly distilled models may lack the safeguards of their teachers~\citep{anthropic2026distillation, GoogleCloud2026_GTIG_AI_Threat_Tracker}. Providers and policymakers who respond to this threat by deploying output perturbation defenses are implicitly adopting a threat model, whether they know it or not.

\vspace{0.5em}

Our results make the stakes of this concrete: a defense that closes most of the distillation gap under one threat model can be neutralized by a one-token prefix under another. A coordinated industry and policy response to distillation requires a shared understanding of what existing defenses actually protect against and under what assumptions. Our framework provides the vocabulary for that conversation, by requiring that every effectiveness claim be tied to a specific $(Q, D, P)$ tuple, making the assumed attacker explicit and contestable rather than obscured and unreported.

\vspace{-1em}

\paragraph{Choosing the threat model in practice.}
One remaining question is which threat model to choose. Our framework suggests three principles for calibrating that choice. 

First, budgets should be grounded in the attacker's realistic capabilities rather than the conventions of the evaluation setup. The query and data budgets should reflect what an attacker could afford relative to their next-best alternative, such as training a comparable model from scratch, rather than to the size of the benchmark dataset. $Q$ can also be calibrated to legitimate user behavior. Our matched-utility budget (Section~\ref{sec:cost_analysis}) is one such example: setting $Q$ equal to what a legitimate user would spend to obtain correct answers under the degraded outputs the defense produces. 

Second, the interface profile should reflect the API behind which the defense will actually be deployed. Since APIs evolve, a defense should be re-evaluated whenever its interface profile changes. 

Third, papers whose primary contribution is a defense should report results across multiple points in the threat-model space rather than a single implicit baseline. This mirrors established conventions in adversarial machine learning, where defenses are expected to be evaluated against adaptive adversaries and across a range of attack strengths rather than a single fixed setting~\citep{carlini2019evaluating, honigadversarial, qievaluating, nasr2025attacker}. Adopting the same standard for output perturbation distillation defenses would allow readers and deployers to assess robustness across the range of adversaries they may actually face.
\vspace{-1em}

\section*{Impact Statement}
This paper aims to highlight that proposed defenses against distillation can be effective under certain threat models but not under others, and that evaluating them without an explicit threat model can lead to an overestimation of their effectiveness. The attacks used to demonstrate the framework rely solely on techniques documented in previous work, and do not introduce new attack capabilities.

\bibliography{references}
\bibliographystyle{icml2026}

\newpage
\appendix
\onecolumn
\section{Antidistillation Sampling} \label{app:ads_details}

\paragraph{Mechanism.}
Antidistillation Sampling (ADS) is a decoding-time defense that perturbs the teacher model's next-token distribution to reduce the effectiveness of model distillation by biasing sampling away from tokens that are informative to a student model.

 Given a prompt $x$ and partial generation $y_{<t}$, let $p_\theta(y_t \mid x, y_{<t})$ denote the teacher model's next-token distribution. Standard sampling draws $y_t \sim p_\theta$. ADS instead samples from a perturbed distribution $\tilde{p}_\theta$ defined as:
\begin{equation}
\tilde{p}_\theta(y_t \mid x, y_{<t}) \propto \exp\left(\frac{1}{\tau} \log p_\theta(y_t \mid x, y_{<t}) + \lambda \cdot \widehat{\Delta}(y_t \mid x, y_{<t})\right)
\end{equation}
where $\tau$ is the temperature, $\lambda \geq 0$ controls the perturbation strength, and $\widehat{\Delta}(y_t \mid x, y_{<t})$ is an antidistillation penalty term.

The penalty $\widehat{\Delta}$ is computed using a proxy student model $\phi$ and its loss gradient $g = \nabla \ell(\phi)$. It approximates the directional derivative of the proxy's next-token log probabilities along $g$, which captures the rate at which the downstream loss would increase if the proxy were updated to favor token $y_t$:
\begin{equation}
\widehat{\Delta}(y_t \mid x, y_{<t}) = \frac{\log p_{\phi + \epsilon g}(y_t \mid x, y_{<t}) - \log p_{\phi - \epsilon g}(y_t \mid x, y_{<t})}{2\epsilon}
\end{equation}
where $\epsilon$ is a small perturbation parameter and $g$ is precomputed once. This requires two forward passes through the proxy model per generation step, evaluating at $\phi \pm \epsilon g$. Tokens with large positive $\widehat{\Delta}$ are those that would degrade the proxy's downstream performance if the student were trained on them. ADS up-weights these tokens in the sampling distribution, biasing the teacher toward outputs that are less informative for distillation.

The parameter $\lambda$ controls the utility-distillability trade-off. At $\lambda = 0$, ADS reduces to standard temperature sampling. As $\lambda$ increases, the teacher's output quality degrades while student distillation performance degrades faster. For ADS to be practically useful, the teacher must retain high utility (i.e., high accuracy). We set this boundary at 70\% teacher accuracy, corresponding to a maximum absolute drop of 20\%. Using the original paper's setup, this boundary corresponds to $\lambda \leq 0.107$. We call this the \emph{high-utility regime} and focus our analysis on it.

\begin{figure}
    \centering
    \includegraphics[width=0.9\linewidth]{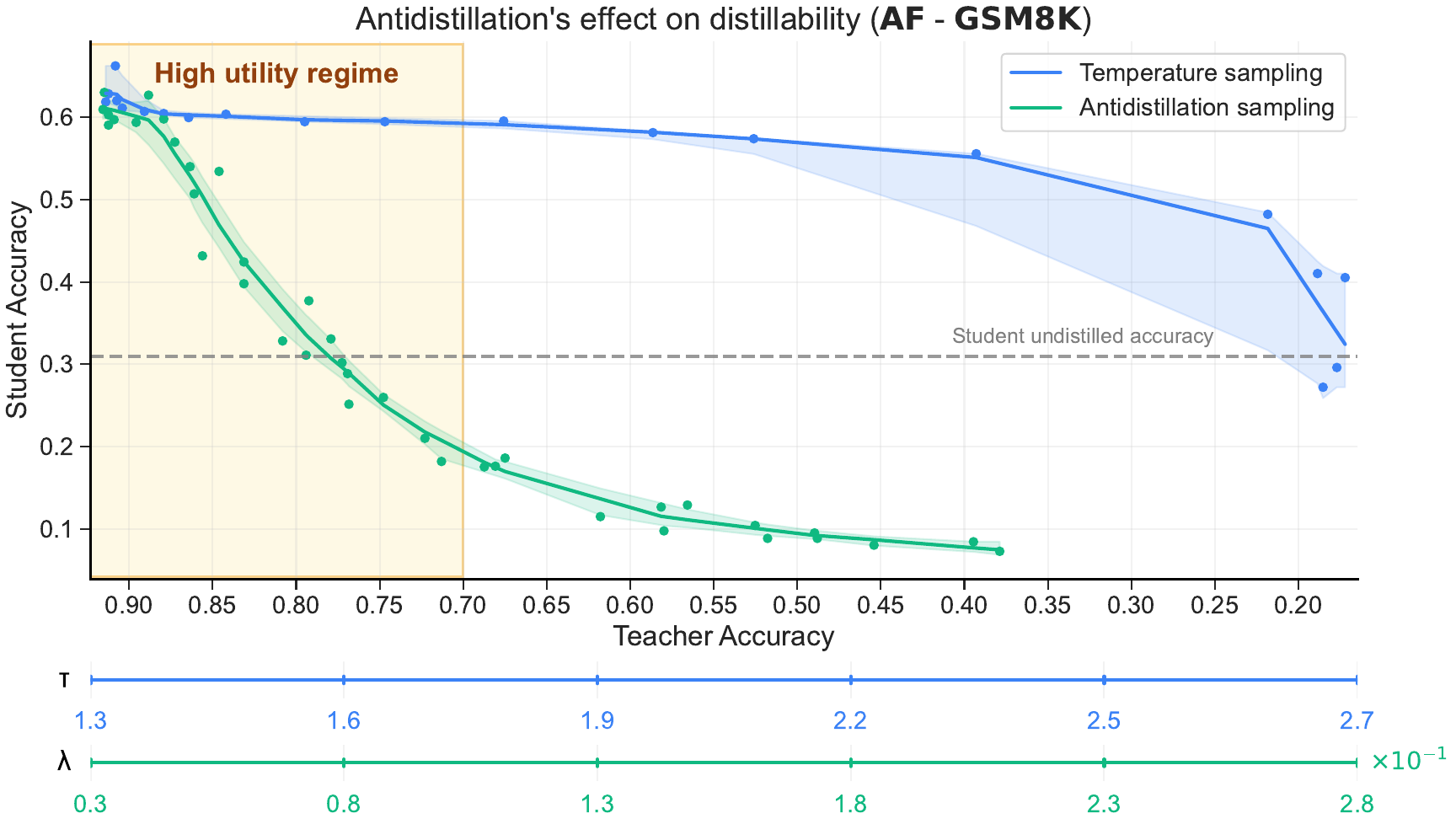}
    \caption{\textbf{Under Antidistillation Sampling (ADS), small reductions in teacher accuracy lead to disproportionately large student performance degradation.} On GSM8K (Answer Forcing evaluation), in the high-utility regime (teacher accuracy $\geq 0.7$), an approximately $20\%$ teacher accuracy reduction corresponds to a $40\%$ student accuracy degradation, whereas temperature sampling produces negligible reduction in student accuracy across the same range.}
    \label{fig:ads_setup}
\end{figure}

This regime is highlighted in Figure~\ref{fig:ads_setup}. At equivalent teacher performance levels, ADS (green curve) achieves substantially lower student accuracy than temperature sampling (blue curve). When teacher accuracy decreases by approximately 20\%, temperature sampling shows only a slight reduction in student accuracy, while ADS reduces student accuracy by approximately 40\%, even dropping below the undistilled baseline.

\paragraph{Noisiness of ADS Generations.} ADS outputs frequently contain recurring noisy artifacts. At structural boundaries, such as the start of the reasoning trace and paragraph transitions, the perturbation often injects semantically irrelevant, high-perplexity tokens that disrupt the coherent chain-of-thought structure. ADS also tends to produce long repetitive sequences of nonsensical tokens: once repetition begins, continuing the pattern becomes increasingly likely under the autoregressive model, often leading to continued repetition until the maximum generation length is reached, without producing further meaningful output. Concrete examples of antidistilled generations are provided in Appendix~\ref{app:example_traces}. 

\section{Extended Results}\label{app:extended_results}
\subsection{Output Post-Processing} \label{sec:app_output_post_processing}
We assume black-box access, with query budget equal to data budget ($Q = D = 5{,}231$) as our threat model. As in previous work, we restrict each prompt to be queried exactly once. Since all local computation is free under our cost model, we evaluate six post-processing strategies applied to the collected ADS outputs before student training:
\paragraph{Repetition Deletion (RD).} RD is applied iteratively for three passes, and in each pass several types of redundancy are removed: character-level repetitions are reduced by collapsing any sequence of five or more identical consecutive characters into a single character, consecutive identical lines are deleted while preserving newline markers (\texttt{\textbackslash n}), and repeated token $n$-grams of up to length $6$ are collapsed to a single occurrence.

\paragraph{RD + Perplexity filtering.} After RD, we filter low-probability tokens using the student model in two passes. In the first pass, we build a context by discarding tokens whose perplexity given the evolving context exceeds a threshold $\tau_1$. In the second pass, we re-evaluate the original sequence using the first-pass context and remove tokens with perplexity above $\tau_2$. The remaining tokens form the final trace.
    
\paragraph{Student-based rephrasing.} We apply few-shot prompting with the student model to clean ADS traces. The prompt includes two examples of noisy traces paired with their cleaned versions, and the model is then used to rephrase new outputs accordingly. Using the student model ensures that no information beyond what is already available to the attacker is introduced into the trace.
    
\paragraph{LLM-based token deletion.} We use an external LLM (GPT‑4o mini) to identify noisy tokens from reasoning traces. Each trace is tokenized and represented as JSON, including each token's index and textual form. The LLM returns the indices of the tokens to delete, while the first boxed answer is masked so that it is preserved. This deletion-only design prevents the external model from injecting new information into the trace, making it safe to use a model different from the student model for that purpose.

\paragraph{Self-talk injection.} Inspired by PART~\citep{ding2025information}, which removes self-talk behaviors as a defense, we insert them as an attack. At paragraph and sentence boundaries, and immediately before reasoning transitions (\emph{therefore, thus, so, then, next, first, finally}, =, \texttt{\textbackslash n}), we insert phrases sampled uniformly from \{\emph{hmm, wait, okay, let me think, well, alright, right, so}\} with probability $0.6$, capping insertions at $12$ per trace.

\paragraph{Incorrect-trace filtering.} We discard traces whose final answers are incorrect.

\paragraph{Results.}
\begin{figure}
    \centering
    \includegraphics[width=0.9\linewidth]{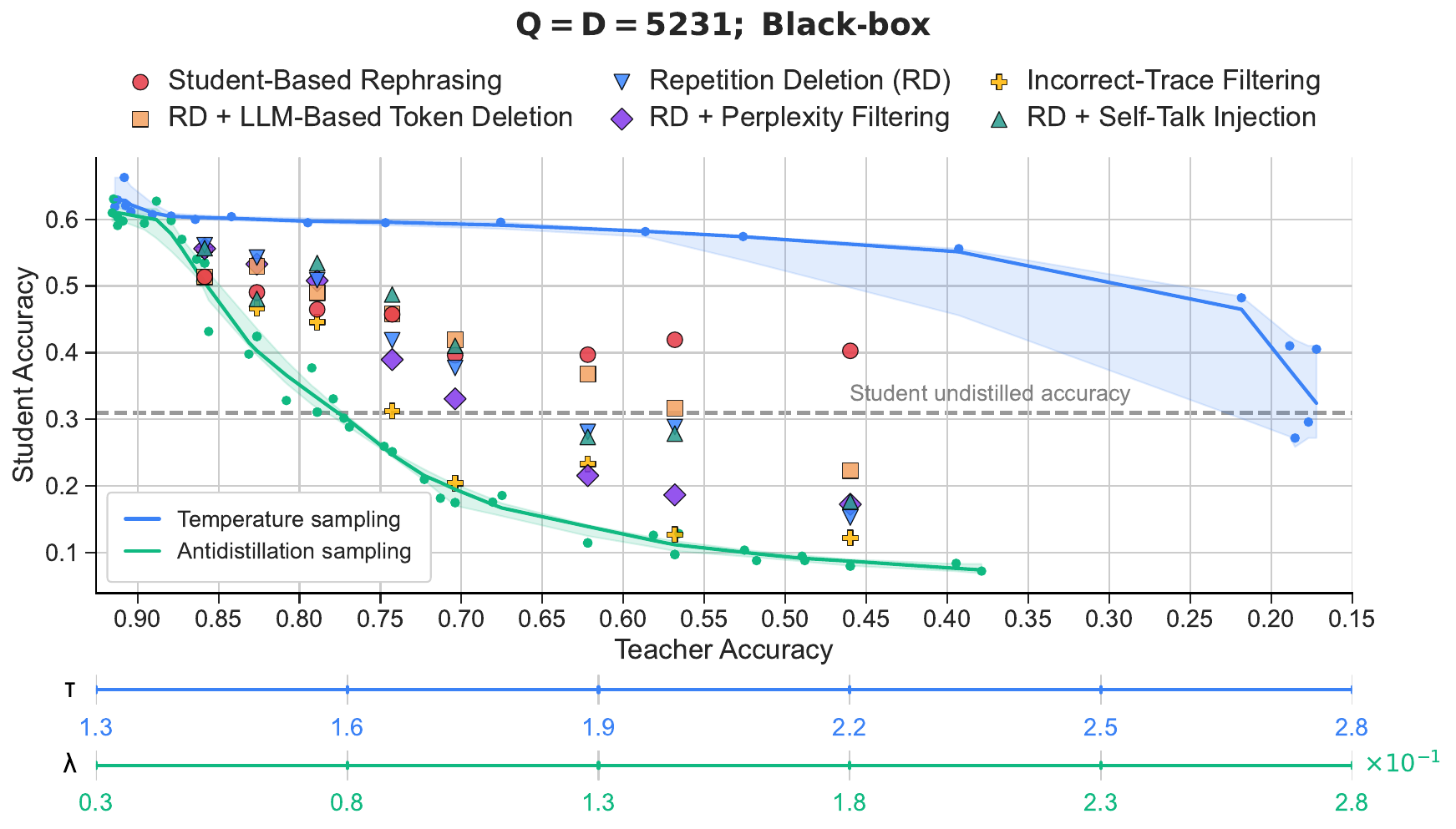}
    \caption{\textbf{Comparison of post-processing strategies against ADS.} Student accuracy on GSM8K across the full teacher accuracy range. The threat model here is $Q=D=5231$, black-box access, and each prompt is queried exactly once. Post-processing strategies considerably weaken the effect of ADS, particularly in the high-utility regime.}
    \label{fig:postprocessing_comparison}
\end{figure}
Figure~\ref{fig:postprocessing_comparison} compares these strategies. LLM-based token deletion combined with repetition deletion reduces the effect of ADS to approximately linear degradation: reductions in teacher accuracy translate into roughly equal reductions in student accuracy. Student-based rephrasing shows a similar trend in the high-utility range before remaining roughly constant at around $0.4$. Self-talk injection combined with repetition deletion offers little improvement over repetition deletion alone, whereas perplexity filtering has a negative effect when combined with repetition deletion. Incorrect-trace filtering produces only marginal improvements. In Section~\ref{sec:attacks}, we use repetition deletion.

\subsection{Weak-Attacker Model}
\begin{figure}
    \centering
    \includegraphics[width=0.9\linewidth]{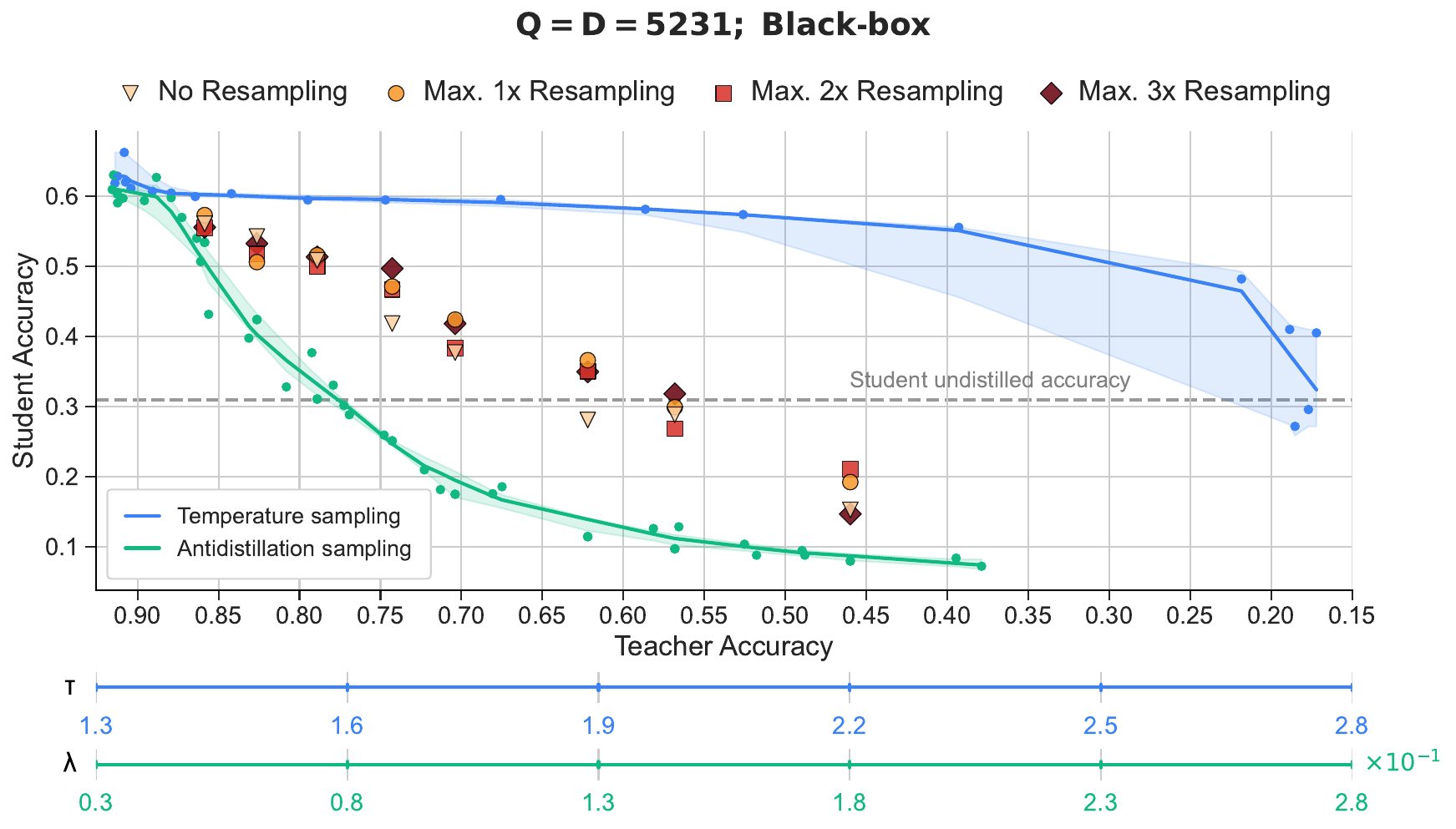}
    \caption{\textbf{Different resampling strategies yield comparable student accuracy.} Student accuracy on GSM8K under black-box max-$k$ resampling ($k \in \{1,2,3\}$) and no resampling, with repetition deletion applied in all cases, across the full $\lambda$ range ($Q = D = 5{,}231$). Differences between resampling strategies remain small across all $\lambda$ values.}
    \label{fig:query_allocation_app}
\end{figure}
\begin{figure}
    \centering
    \includegraphics[width=0.95\linewidth]{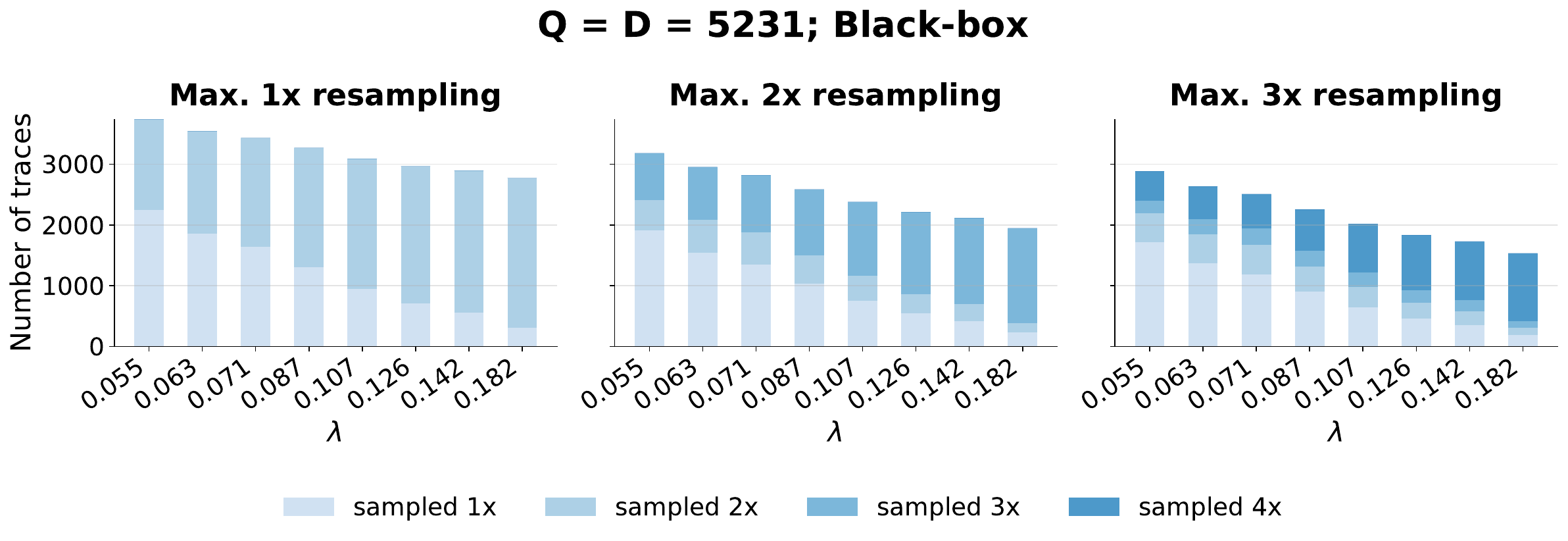}
    \caption{\textbf{Higher $\mathbf{k}$ concentrates more budget on initially incorrect prompts, at the cost of reduced prompt coverage.} Distribution of query budget allocation across prompts under max-$k$ resampling for $k \in \{1,2,3\}$. Bars indicate how many prompts receive $1$ to $k+1$ generations.}
    \label{fig:resampling_distribution}
\end{figure}

We consider the same black-box threat model as in the main experiments with a fixed query budget equal to the dataset size ($Q = D = 5{,}231$). Figure~\ref{fig:query_allocation_app} extends the main-text results to the full $\lambda$ range. The differences between resampling strategies remain small throughout. Figure~\ref{fig:resampling_distribution} shows how the query budget is distributed across prompts in each strategy. The height of each bar indicates the total number of prompts queried at least once (i.e., the number of traces in the student training dataset), while the stacked colors within each bar show how that budget is allocated: specifically, how many prompts received exactly $1, 2, \dots, k + 1$ generations in total. Increasing $k$ concentrates more budget on initially incorrect prompts at the expense of prompt coverage.

\subsection{Scaling the Query Budget}
\begin{figure}
    \centering
    \includegraphics[width=0.85\linewidth]{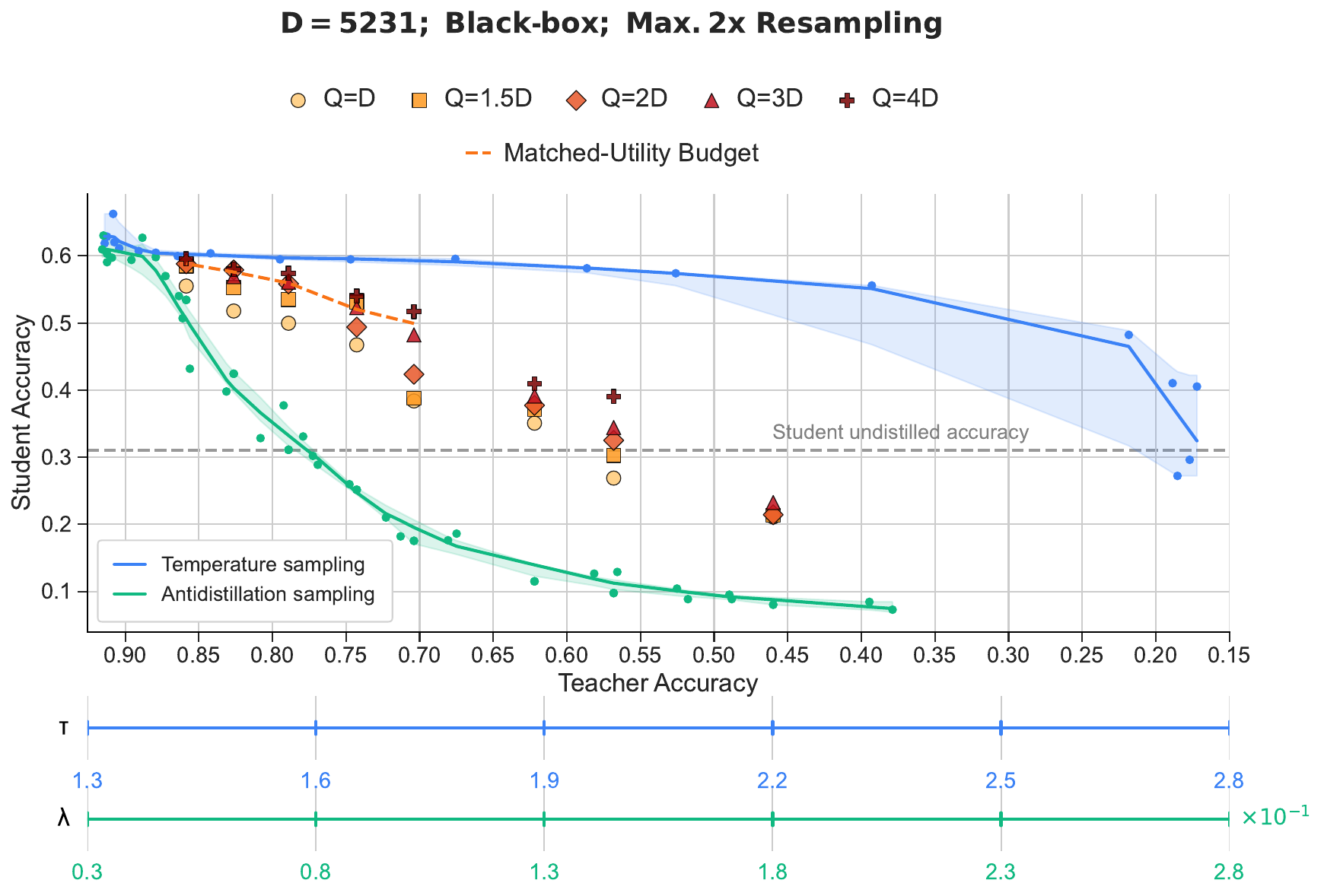}
    \caption{\textbf{Increasing query budget improves student accuracy, with smaller gains under stronger perturbation.} Student accuracy under max-$2$ resampling with repetition deletion, for varying query budgets $Q \in \{D, 1.5D, 2D, 3D, 4D\}$ across the full $\lambda$ range under black-box access. The matched-utility budget is shown only for $\lambda$ values where at least $90\%$ of prompts are solved within $16$ sampling attempts. At high $\lambda$, differences across query budgets become negligible, as perturbation prevents resampling from leading to useful traces.}
    \label{fig:budget_allocation_2}
\end{figure}

\paragraph{Max-2 resampling.}  Figure~\ref{fig:budget_allocation_2} extends the main-text results to the full $\lambda$ range. Student accuracy increases consistently with $Q$ across most $\lambda$ values, confirming the trend reported in the main text. For the largest $\lambda$ value, differences across query budgets become negligible, as the ADS perturbation is strong enough that resampling rarely recovers correct or clean traces regardless of budget. We report the matched-utility budget only for $\lambda \leq 0.107$, where at least $90\%$ of prompts are solved within $16$ sampling attempts, ensuring that the empirical mean is based on the vast majority of cases and remains a meaningful lower bound. For larger $\lambda$ values, this condition fails. In particular, at $\lambda = 0.182$ only $67\%$ of the prompts are solved within $16$ attempts, so the average requery count cannot be reliably estimated from our data. Crucially, even the conservative estimates that we can compute (see Figure~\ref{fig:analysis_mean_steps}) already exceed the highest query budget shown, implying that the matched-utility budget would lie beyond $Q = 4D$. 

\begin{sidewaysfigure}[p]
    \centering
    \begin{subfigure}{\textwidth}
        \centering
        \includegraphics[width=\linewidth]{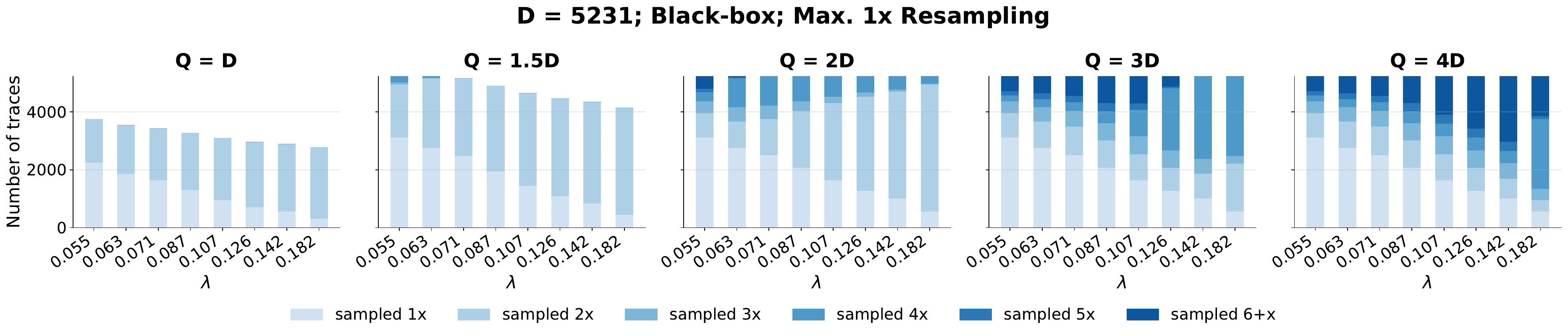}
    \end{subfigure}
    \vspace{1em}
    \noindent\rule{\textwidth}{0pt}
    \begin{subfigure}{\textwidth}
        \centering
        \includegraphics[width=\linewidth]{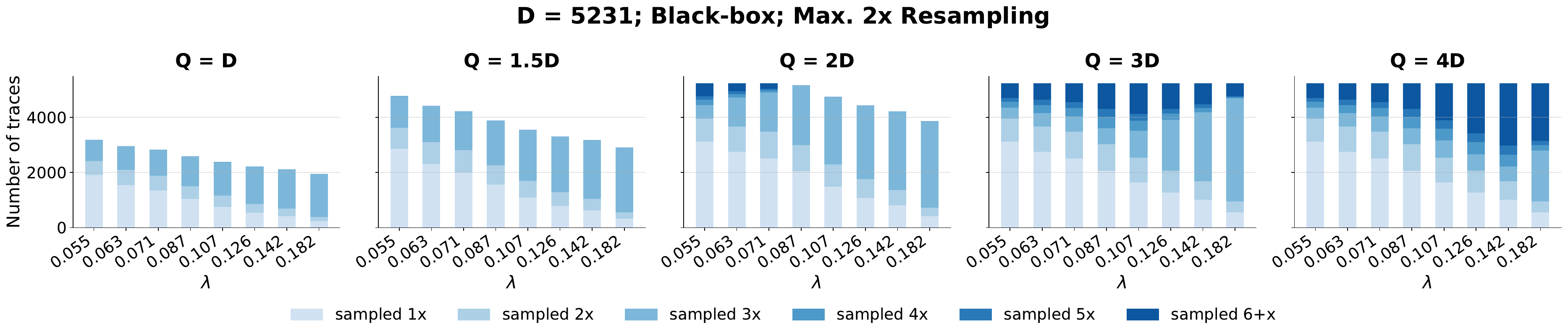}
       
    \end{subfigure}
    \vspace{1em}
    \noindent\rule{\textwidth}{0pt}
    \begin{subfigure}{\textwidth}
        \centering
        \includegraphics[width=\linewidth]{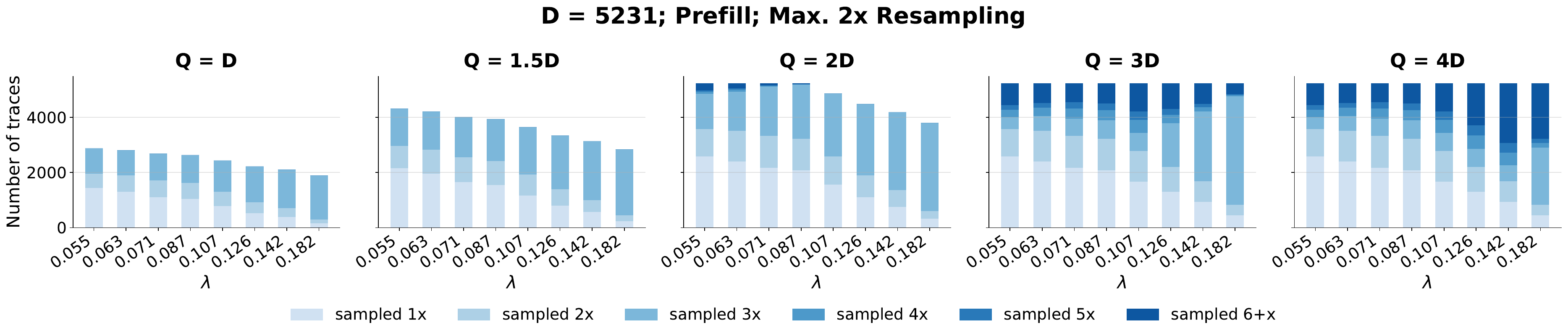}
       
    \end{subfigure}
    \caption{\textbf{Query budget distribution across prompts.} Per-prompt generation counts under max-$1$ resampling (\textbf{top}) and max-$2$ resampling for black-box (\textbf{middle}) and prefill access (\textbf{bottom}), for query budgets $Q \in \{D, 1.5D, 2D, 3D, 4D\}$.}
    \label{fig:strategy_q_both}
\end{sidewaysfigure}

Figure~\ref{fig:strategy_q_both} shows how the query budget is distributed across traces under each value of $Q$. For $Q \geq 3D$, every prompt has been queried at least once and the remaining budget is spent on resampling.

\begin{figure}
    \centering
    \includegraphics[width=0.85\linewidth]{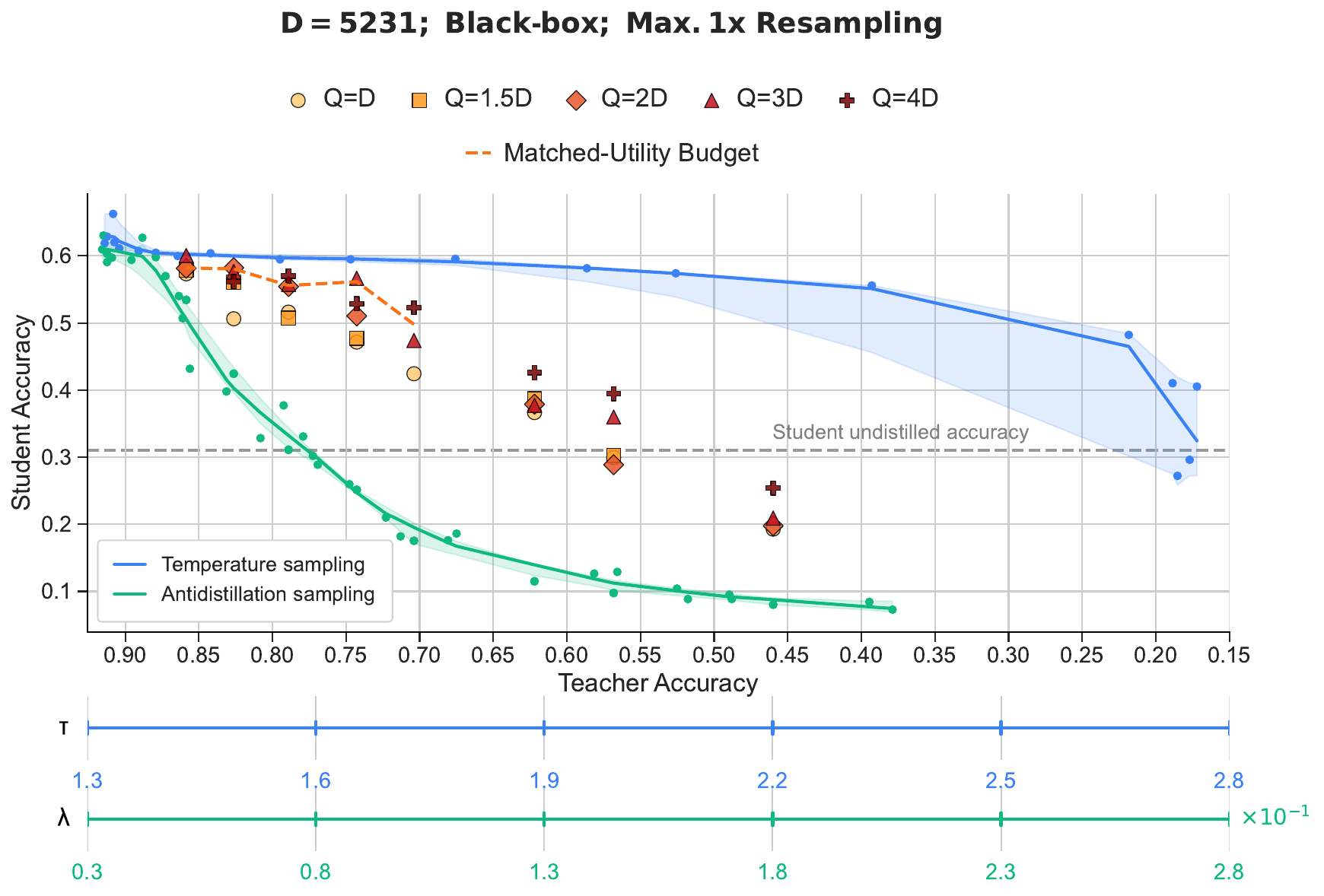}
    \caption{\textbf{Max-$\mathbf{1}$ resampling yields student accuracy nearly identical to max-$\mathbf{2}$.} Student accuracy under max-$1$ resampling with repetition deletion, for varying query budgets $Q \in \{D, 1.5D, 2D, 3D, 4D\}$, across the full $\lambda$ range.}\label{fig:query_allocation_max1}
\end{figure}

\paragraph{Max-1 resampling.} 
Figure~\ref{fig:query_allocation_max1} extends the analysis to max-$1$ resampling across varying query budgets. Student accuracy is very similar to the max-$2$ results, with no meaningful difference between the two strategies at any $\lambda$ value, further confirming that the gain from increasing $Q$ is not sensitive to the exact resampling cap. Figure~\ref{fig:strategy_q_both} shows that under max-$1$ resampling the number of final traces already reaches its maximum at $Q = 2D$, after which all budget is spent on resampling rather than expanding prompt coverage.

\subsection{Interface Profile: Enabling Prefill Access} \label{sec:prefill_access_appendix}
\begin{figure}
    \centering
\includegraphics[width=0.9\linewidth]{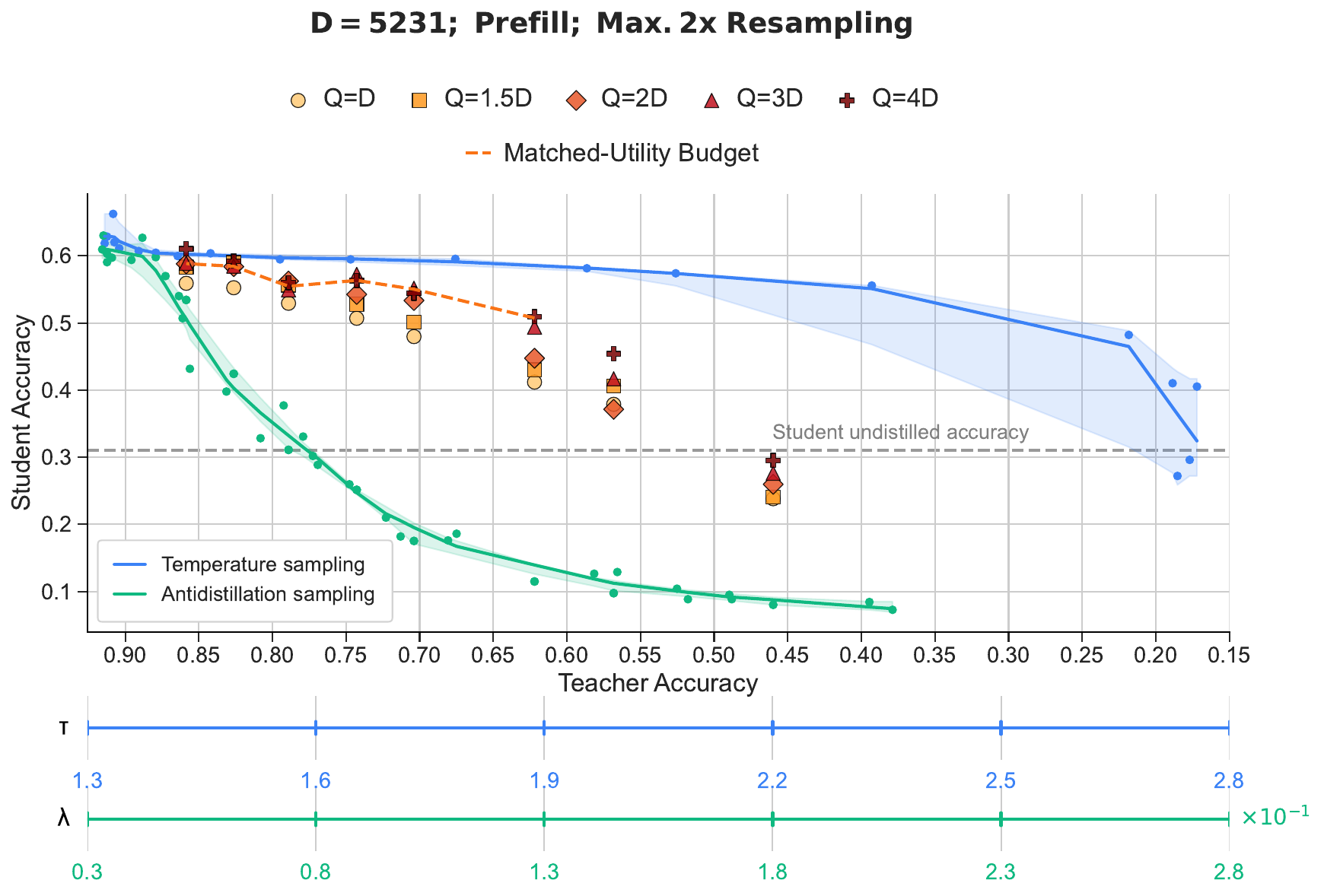}
\caption{\textbf{Prefill access yields consistently higher student accuracy than black-box access (Figure \ref{fig:budget_allocation_2}) at the same query budget.} Student accuracy under max-$2$ resampling with repetition deletion and prefill access (prefix ``First''), for varying query budgets $Q \in \{D, 1.5D, 2D, 3D, 4D\}$, across the full $\lambda$ range. The matched-utility budget is shown for $\lambda$ values where at least $90\%$ of prompts are solved within $16$ resampling attempts.}\label{fig:full_result_first}
\end{figure}

\begin{figure}
    \centering
    \begin{subfigure}[t]{0.325\textwidth}
        \centering
        \includegraphics[width=\linewidth]{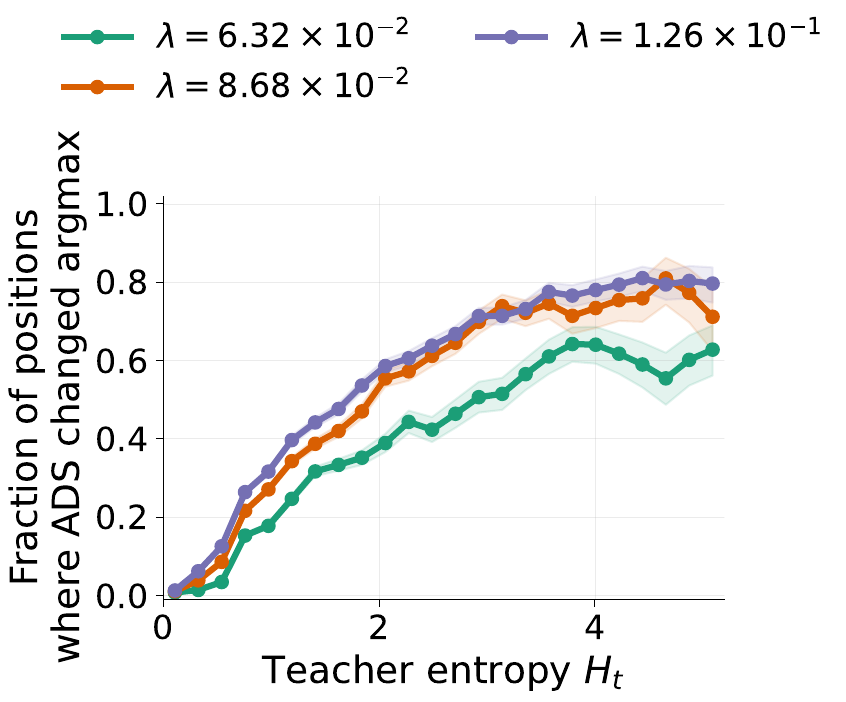}
        \caption{Fraction of positions where ADS changes the argmax token as a function of teacher entropy $H_t$, under black-box access. Higher entropy corresponds to a flatter next-token distribution, reducing the gap between the top token and its competitors and making it easier for ADS's perturbation to flip the argmax.}
        \label{fig:fraction_changed}
    \end{subfigure}
    \hfill
    \begin{subfigure}[t]{0.29\textwidth}
        \centering
        \includegraphics[width=\linewidth]{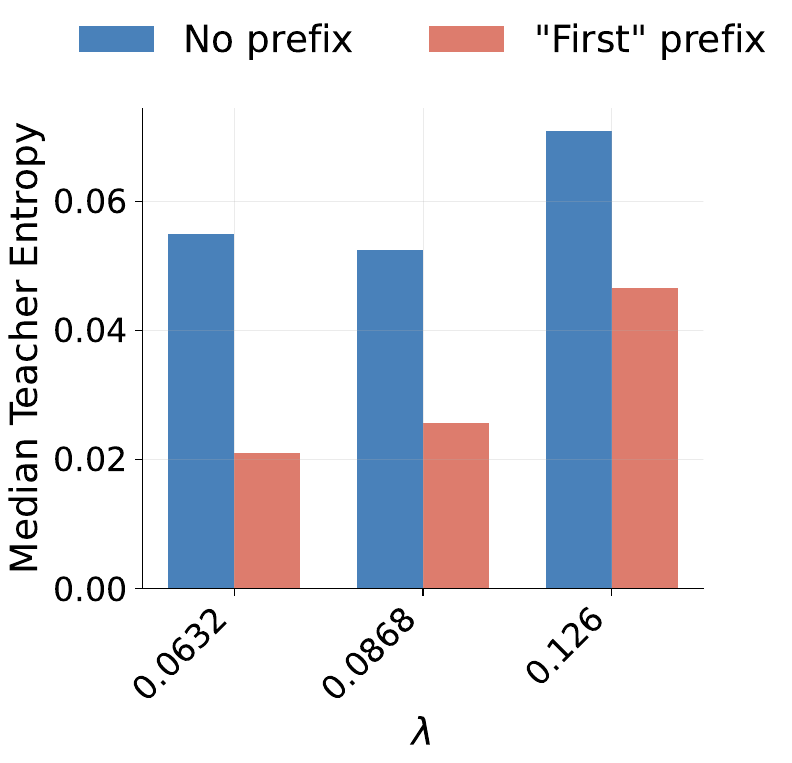}
        \caption{Median teacher entropy across all generated tokens, for each $\lambda$ value, with and without the ``First'' prefix. The prefix consistently lowers entropy, reducing ADS's ability to flip the argmax.}
        \label{fig:median}
    \end{subfigure}
    \hfill
    \begin{subfigure}[t]{0.325\textwidth}
        \centering
        \includegraphics[width=\linewidth]{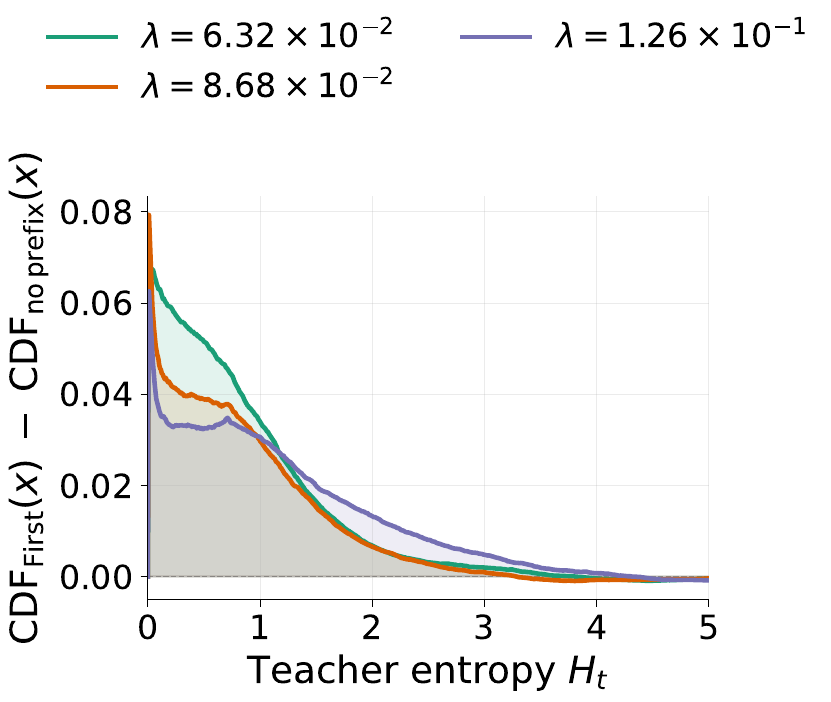}
        \caption{Difference in empirical CDFs as a function of teacher entropy. Positive values indicate the ``First'' prefix shifts probability mass toward lower-entropy distributions, reducing ADS effectiveness.}
        \label{fig:fraction_changed_first}
    \end{subfigure}
    \caption{\textbf{Effect of teacher entropy and prefill access on ADS's ability to change the argmax token.} When the next-token distribution is uncertain (high entropy), ADS can more easily redirect generation toward student-harming tokens. Conditioning on a single prefix token reduces entropy at generated positions, limiting ADS's effective perturbation rate.}
    \label{fig:argmax_analysis}
\end{figure}

Figure~\ref{fig:full_result_first} extends the main-text results to the full $\lambda$ range, confirming that prefill access with the prefix ``First'' yields consistently higher student accuracy across all $\lambda$ values. As in the black-box setting, we report the matched-utility budget only for $\lambda$ values where at least $90\%$ of prompts are solved within $16$ resampling attempts. Here, this threshold is met for one additional point: at $\lambda=0.126$, $93.0\%$ of prompts are solved within $16$ attempts, so we include it in the matched-utility curve. The matched-utility line lies higher than in the black-box setting, meaning that ADS is less effective under prefill access. Figure~\ref{fig:strategy_q_both} shows that the distribution of resampling attempts does not differ significantly from the black-box setting. 

Figure~\ref{fig:argmax_analysis} provides insight into why setting ``First'' as a prefix works. ADS perturbs the teacher's next-token distribution by adding a directional term that up-weights tokens estimated to increase the student's loss. Whether this perturbation actually flips the argmax depends on how concentrated the original distribution is. When the entropy $H_t$ is high, the probability mass is spread across many alternatives, and the gap between the most likely token and its alternatives is small. Even a modest perturbation can flip the argmax. As shown in Figure~\ref{fig:fraction_changed}, the fraction of positions at which ADS changes the argmax increases with $H_t$, indicating that more positions in the trace are actively corrupted when the teacher is uncertain. 

Conditioning on the prefix ``First'' reduces this effect. If the initial token is fixed, the entropy of the teacher's distribution at subsequent positions decreases (see Figures~\ref{fig:median} and~\ref{fig:fraction_changed_first}). Lower entropy implies a larger average gap between the argmax token and its alternatives, making ADS-induced argmax flips less likely. Figure~\ref{fig:fraction_changed_first} confirms this: the fraction of positions at which the argmax changes is systematically lower under prefill access than under black-box access. Consequently, the generated trace remains closer to what the teacher would have produced without ADS, yielding cleaner reasoning steps, more correct traces (see Figures~\ref{fig:resampling_convergence} and~\ref{fig:percentage_traces}), and higher student accuracy.

\subsection{Further Analysis}
\begin{figure}[t]
    \centering
    \includegraphics[width=0.9\textwidth]{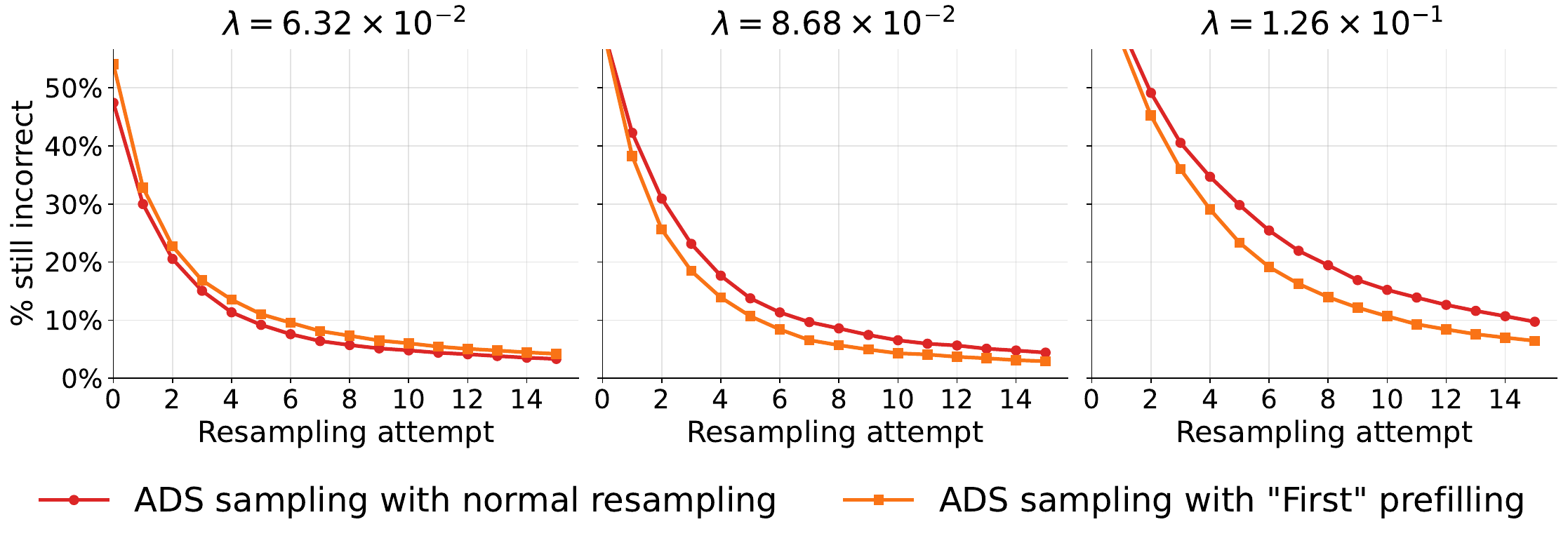}
    \caption{\textbf{Most incorrect traces are corrected within two resampling attempts.} Fraction of incorrect traces remaining across successive resampling attempts, for standard resampling and resampling with prefix ``First'', at three representative $\lambda$ values. Both strategies exhibit exponential-like decay, with most of the reduction occurring within the first two attempts. Prefill access accelerates convergence at high $\lambda$ values but offers little additional benefit at the smallest $\lambda$.}
    \label{fig:resampling_convergence}
\end{figure}

\begin{figure}[t]
    \centering
    \includegraphics[width=0.95\textwidth]{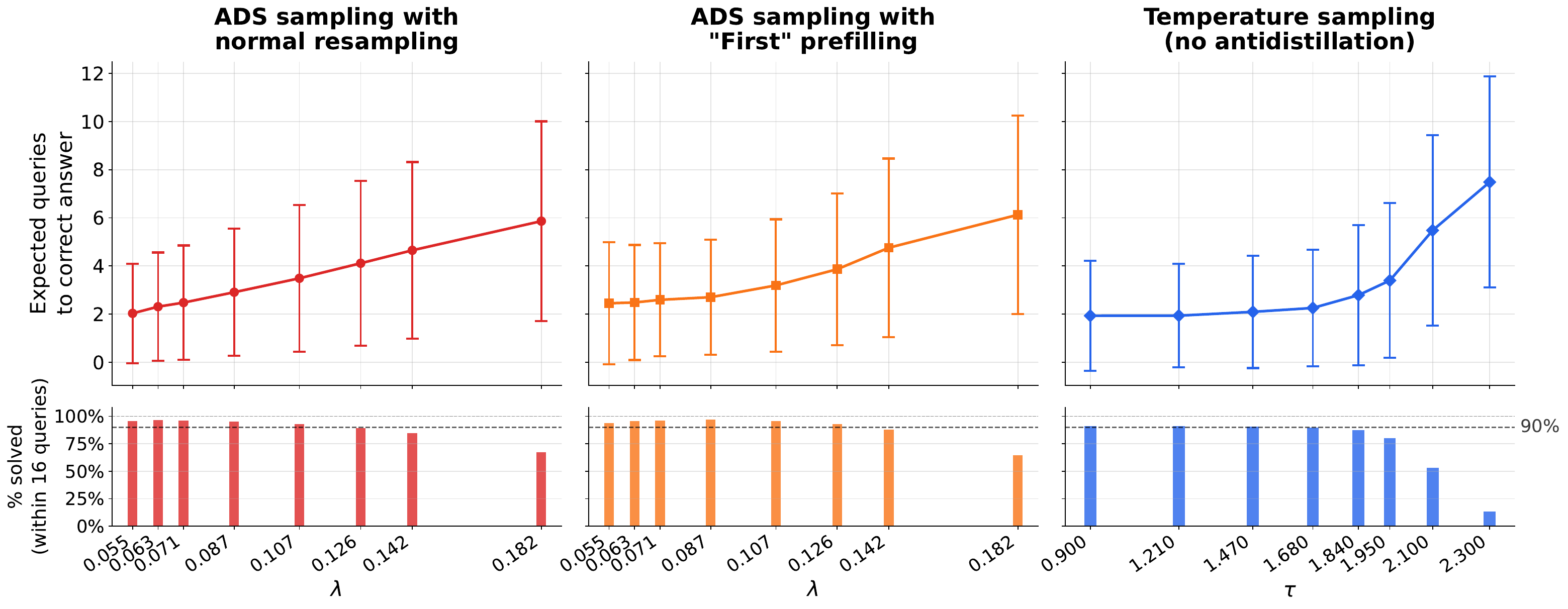}
    \caption{\textbf{Expected queries to a correct answer grow monotonically with perturbation strength.} Expected number of queries to a correct answer as a function of $\lambda$ for ADS (standard black-box resampling and prefill resampling with prefix ``First''), and as a function of $\tau$ for temperature sampling. Error bars denote $\pm 1$ standard deviation. As $\lambda$ increases, the expected number of queries grows monotonically for all three methods. Only traces that achieve a correct answer within 16 queries are included in this estimate (bottom panels show the fraction of such traces). Since this fraction decreases substantially with $\lambda$ and $\tau$, the reported means are computed over an increasingly selective subset and therefore underestimate the true expected query count.}
    \label{fig:analysis_mean_steps}
\end{figure}

Figure~\ref{fig:resampling_convergence} shows how the number of incorrect traces decreases across successive resampling attempts, comparing standard resampling against resampling with the prefix ``First''. Both strategies exhibit exponential-like decay and most of the reduction occurs within the first two attempts, which is why we focus on max-$1$, max-$2$, and max-$3$ resampling in the main text and do not consider higher values. Response prefilling leads to faster decay at moderate and large $\lambda$ values, while offering little additional benefit at the smallest $\lambda$, where most traces are still very clean.

Figure~\ref{fig:analysis_mean_steps} reports the expected number of queries needed to obtain a correct answer as a function of $\lambda$ for ADS and $\tau$ for temperature sampling. The expected query count grows monotonically with $\lambda$ and $\tau$. As discussed above, these estimates are based only on prompts solved within $16$ attempts and therefore underestimate the true expected query count, increasingly so at larger $\lambda$ and $\tau$ values.

\begin{figure}[t]
    \centering
\includegraphics[width=0.95\linewidth]{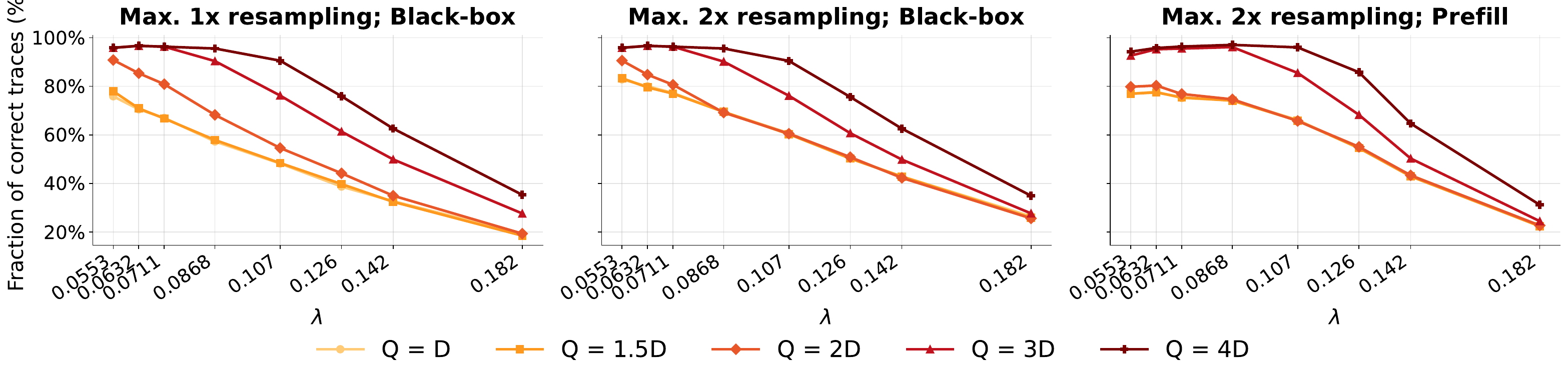}
\caption{\textbf{Increasing query budget recovers more correct 
traces, with the largest gains at moderate $\mathbf{\lambda}$.} Fraction of correct traces retained after max-$2$ resampling, comparing black-box and prefill access across query budgets $Q \in \{D, 1.5D, 2D, 3D, 4D\}$ and $\lambda$ values. The effect of increasing $Q$ is biggest at moderate $\lambda$ values, where ADS's perturbation is neither too weak nor too strong for resampling to make a meaningful difference. Prefill access further increases the fraction of correct traces, particularly at higher query budgets.}\label{fig:percentage_traces}
\end{figure}

Figure~\ref{fig:percentage_traces} shows that increasing the query budget leads to a higher fraction of correct traces in the training dataset. The effect is largest at moderate $\lambda$ values, where the ADS perturbation is neither too weak nor too strong for resampling to make a difference. Prefill access further increases the fraction of correct traces at moderate $\lambda$ values, particularly for $Q \geq 3D$.

\subsection{Effect of Resampling under Temperature Sampling} \label{app:resampling_temperature_buget}
\begin{figure}[t]
    \centering
\includegraphics[width=0.55\linewidth]{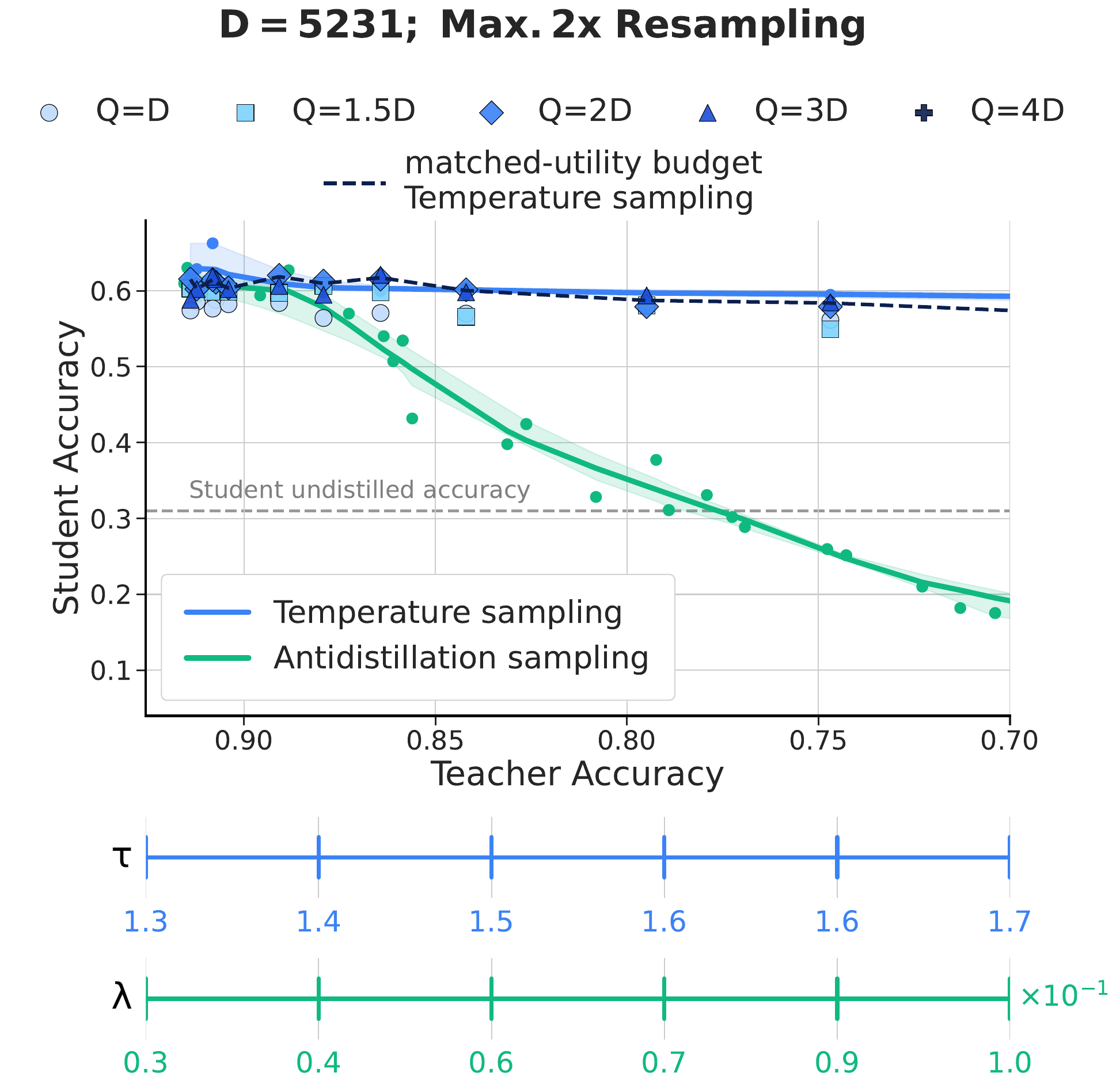}
\caption{\textbf{Resampling has negligible effect under temperature sampling.} Student accuracy under max-$2$ resampling with repetition deletion for varying query budgets $Q \in \{D, 1.5D, 2D, 3D, 4D\}$ under temperature sampling. Unlike ADS, the matched-utility curve is nearly identical to the standard temperature-sampling curve.}\label{fig:resampling_temp}
\end{figure}

For comparison, we also plot the matched-utility budget curve for temperature sampling in Figure~\ref{fig:resampling_temp}, again with repetition deletion and max-2 resampling applied. Unlike the ADS matched-utility curve, the matched-utility curve is almost identical to the standard temperature-sampling curve. This suggests that resampling has a much smaller effect under temperature sampling than under ADS.

\clearpage
\section{Example of Attacked ADS Traces}\label{app:example_traces}
\textbf{Repetition Deletion (RD)}

\includegraphics[width=0.91\textwidth]{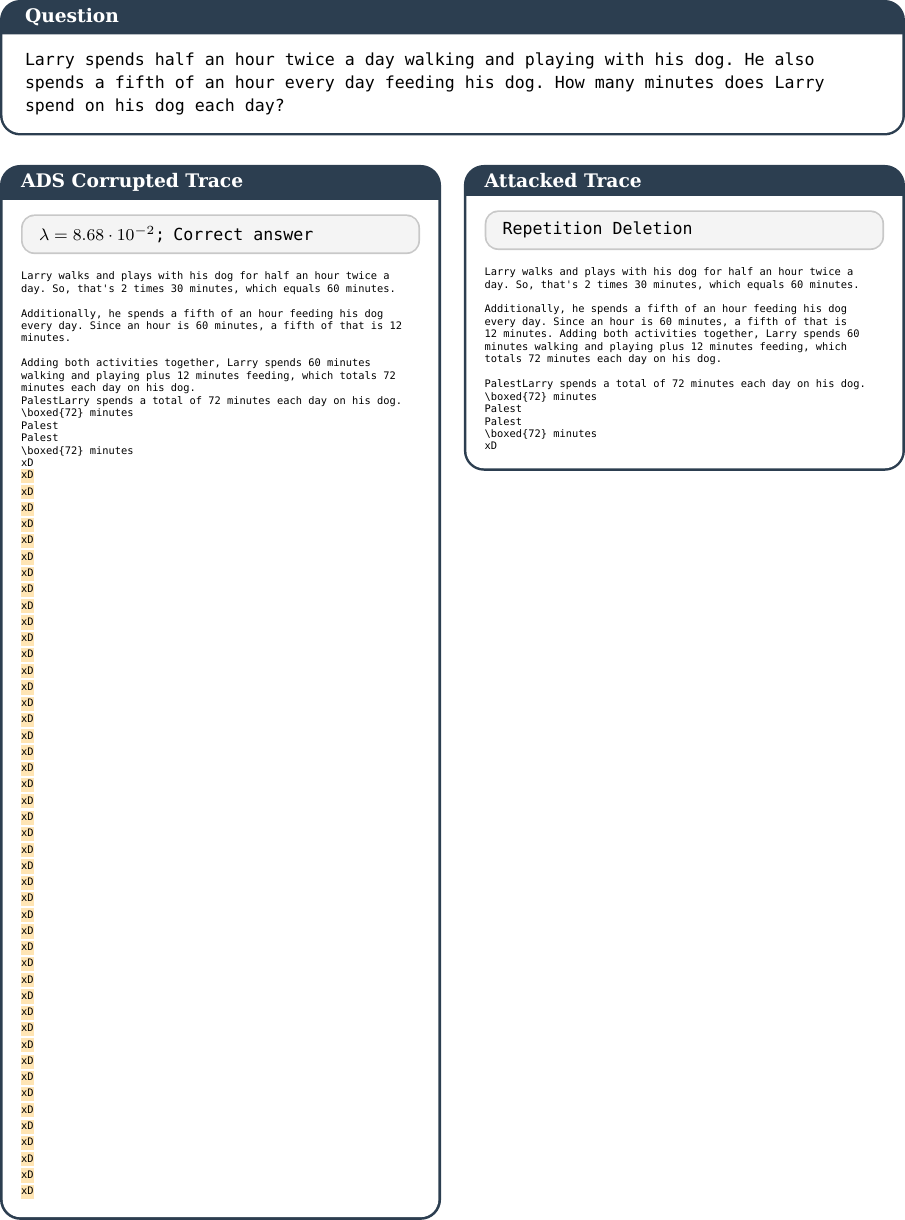}

\clearpage
\textbf{Repetition Deletion + Perplexity Filtering}

\includegraphics[width=0.91\textwidth]{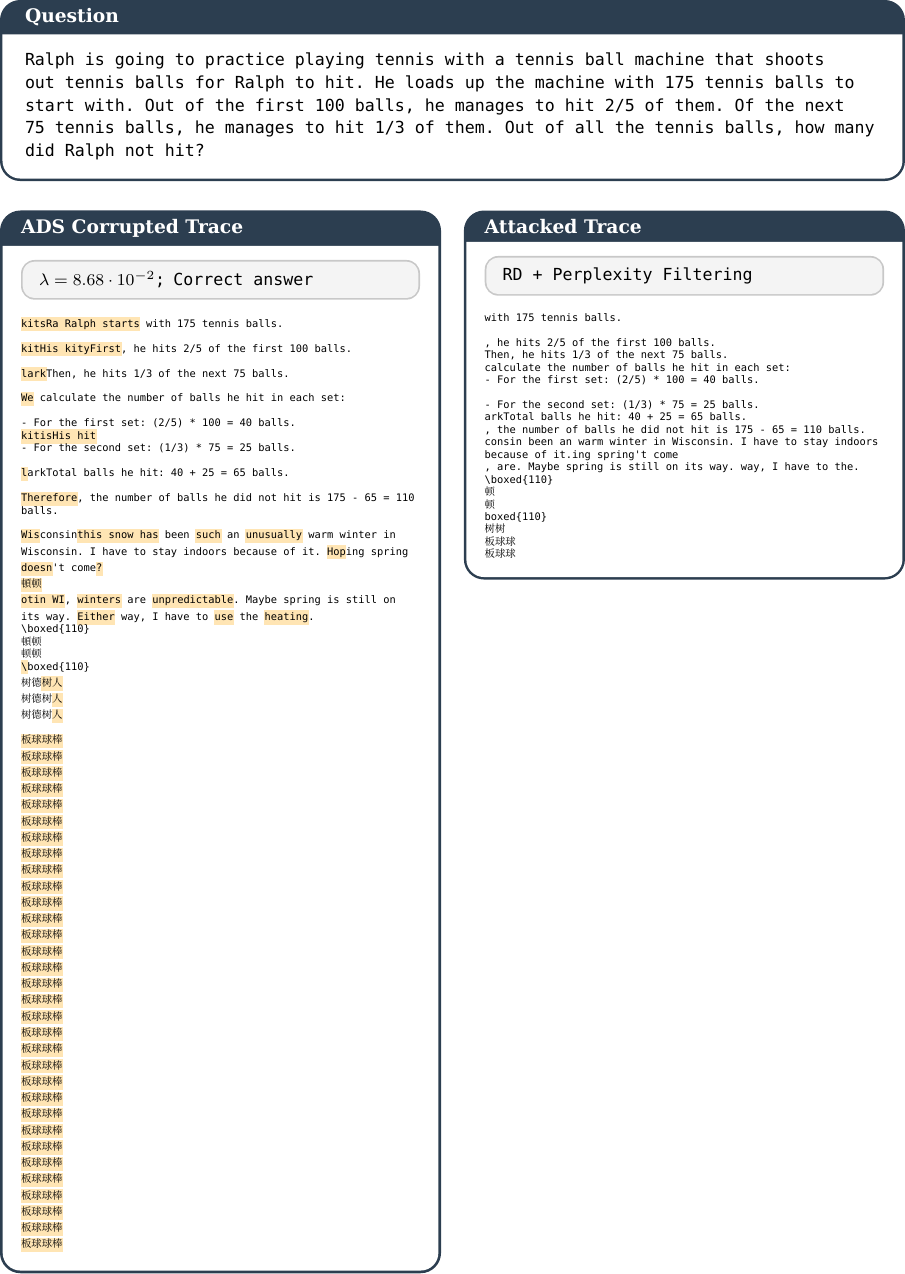}
\clearpage
\textbf{Student-Based Rephrasing}

\includegraphics[width=0.91\textwidth]{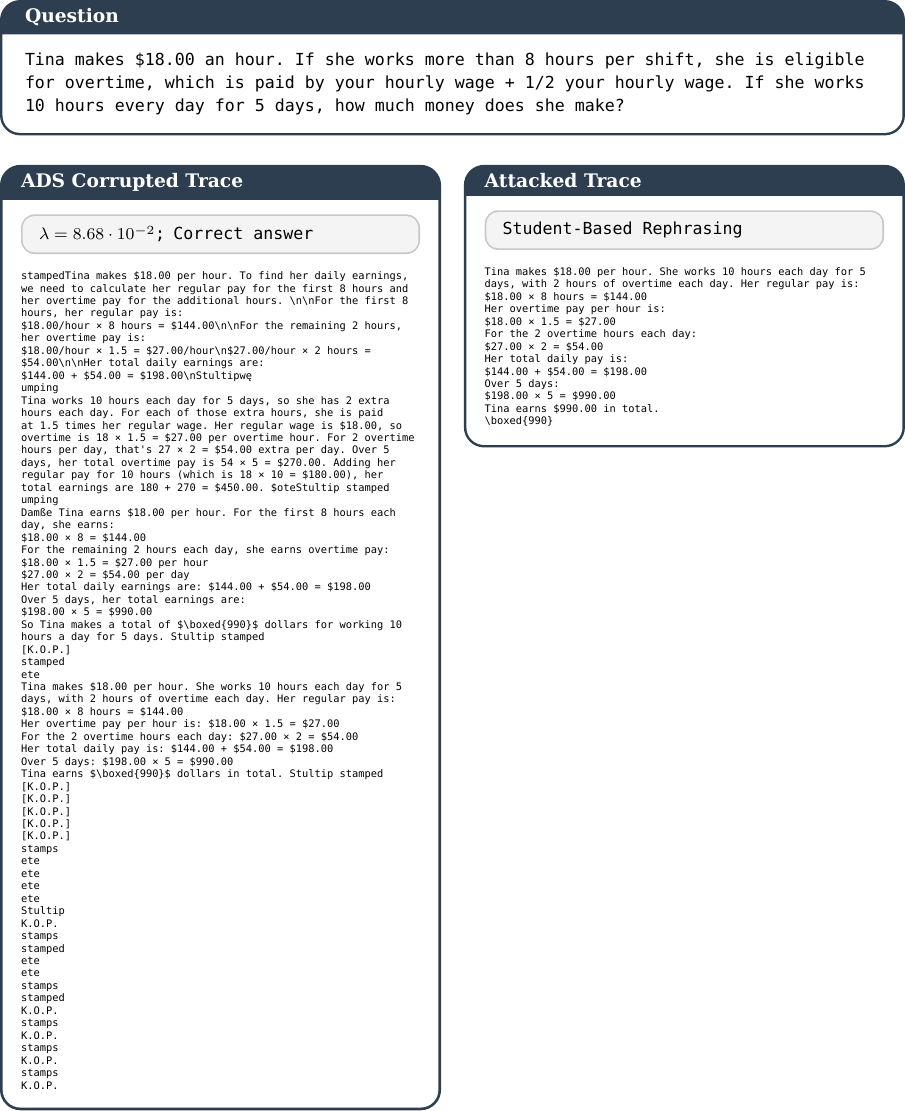}
\clearpage
\textbf{Repetition Deletion + LLM-Based Token Deletion}

\includegraphics[width=0.91\textwidth]{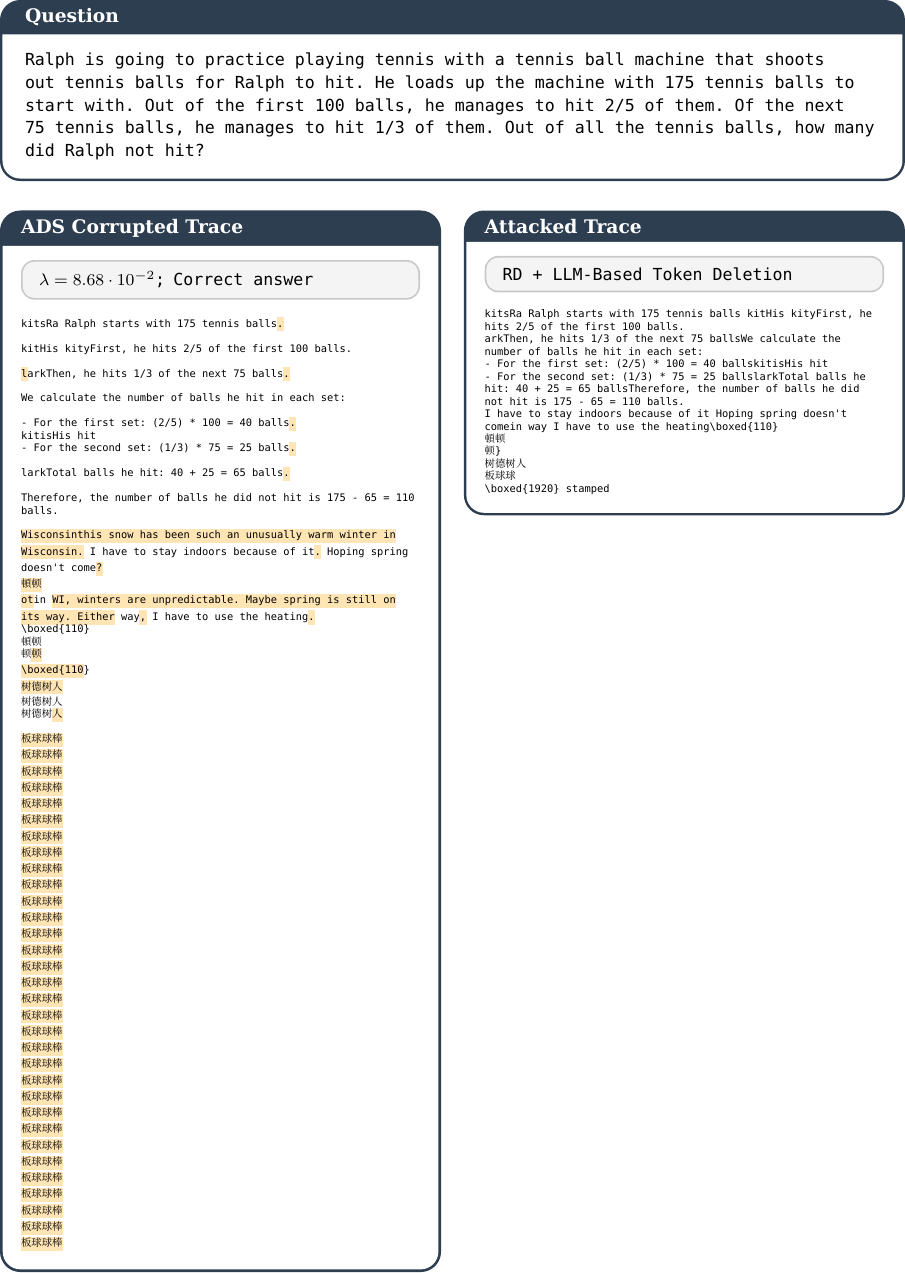}

\clearpage
\textbf{Repetition Deletion + Self-Talk Token Injection}

\includegraphics[width=0.91\textwidth]{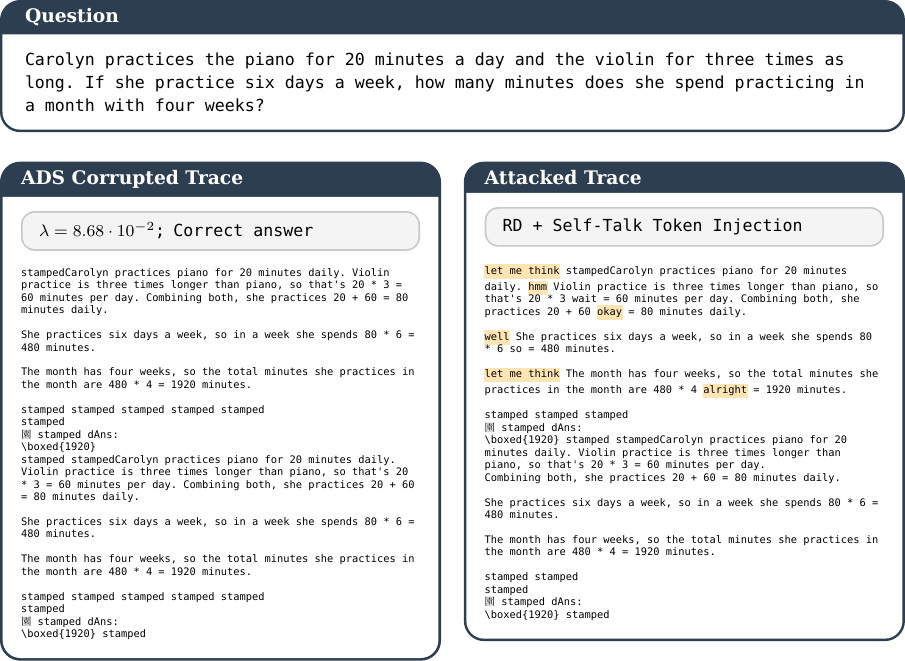}
\clearpage
\clearpage
\textbf{Fixed-Start Resampling + Repetition Deletion}

\includegraphics[width=0.91\textwidth]{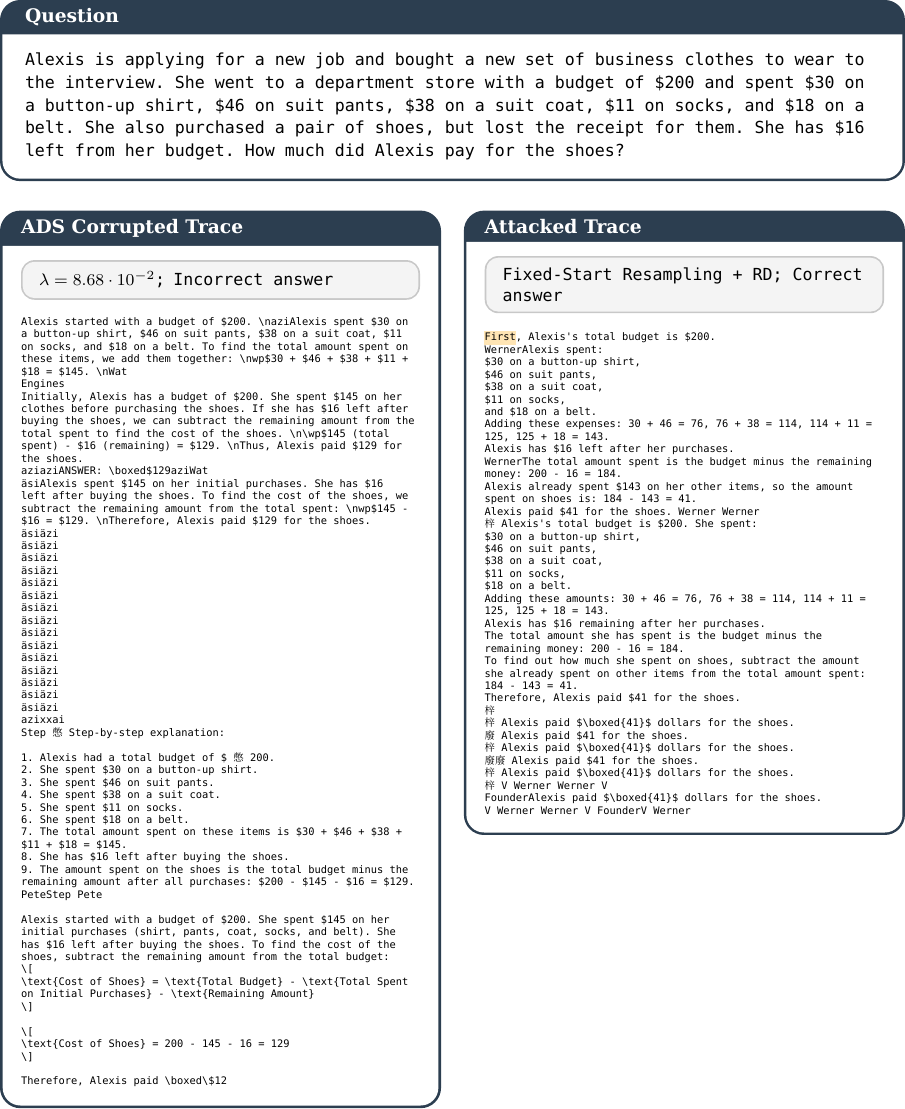}
\clearpage

\section{Notes on the Graphs}
We follow \citet{savani2025antidistillation} in constructing the visualizations. Specifically, we report the mean and 95\% confidence intervals computed over bootstrapped LOWESS fits. For the additional $\tau$ and $\lambda$ axes shown in the figures, we use linear regression, i.e., we fit $\lambda = \beta_0 + \beta_1 \cdot \text{Teacher Accuracy}$ on the current set of data points and then use the fitted coefficients to predict $\lambda$ values from Teacher Accuracy.

The $\lambda$ and $\tau$ axis values in Figure~\ref{fig:ads_setup} differ from those reported in the ADS paper. We attribute this discrepancy to the use of different data points in fitting the linear model.

\end{document}